\documentclass[preprint, superscriptaddress]{revtex4-1}

\usepackage{graphicx}
\usepackage{dcolumn}
\usepackage{amsmath}
\usepackage{amsfonts}
\usepackage{mathrsfs}
\usepackage{amssymb}
\usepackage{physics}
\usepackage{mathtools}
\usepackage{stmaryrd} 
\usepackage{xcolor}
\usepackage{tcolorbox}
\definecolor{moynihan}{RGB}{204,51,51}
\usepackage{hyperref}
\hypersetup{
	colorlinks=true,
	linkcolor=moynihan,
	filecolor=moynihan,      
	urlcolor=moynihan,
	citecolor=moynihan,
	pdfstartview={XYZ null null 1.07}
}
\usepackage[T1]{fontenc}
\usepackage[latin1]{inputenc}
\usepackage{booktabs}
\usepackage[font=small,labelfont=bf,tableposition=top]{caption}
\usepackage{rotating}
\usepackage{array,multirow}
\usepackage{float}
\usepackage{amsmath}
\usepackage{algorithm}
\usepackage{algpseudocode}

\usepackage{makecell} 

\begin{document}
 \begin{center}
\hrule height 4.1pt
\vspace{0.5cm}
{\Large\textbf{Classical-to-Quantum Sequence Encoding in Genomics}}
\vspace{0.5cm}
\hrule height 1.2pt
\end{center}
\author{Nouhaila Innan}\email[]{nouhailainnan@gmail.com}
\affiliation{Quantum Physics and Magnetism Team, LPMC, Faculty of Sciences Ben
M'sick,\\
Hassan II University of Casablanca,
Morocco.}
\author{Muhammad Al-Zafar Khan}\email[]{muhammadalzafark@gmail.com}
\affiliation{Robotics, Autonomous Intelligence, and Learning Laboratory (RAIL), School of Computer Science and Applied Mathematics, University of the Witwatersrand, 1 Jan Smuts Ave, Braamfontein, Johannesburg 2000, Gauteng, South Africa}
\begin{abstract}
\footnotesize{
DNA sequencing allows for the determination of the genetic code of an organism, and therefore is an indispensable tool that has applications in Medicine, Life Sciences, Evolutionary Biology, Food Sciences and Technology, and Agriculture. In this paper, we present several novel methods of performing classical-to-quantum data encoding inspired by various mathematical fields, and we demonstrate these ideas within Bioinformatics. In particular, we introduce algorithms that draw inspiration from diverse fields such as Electrical and Electronic Engineering, Information Theory, Differential Geometry, and Neural Network architectures. We provide a complete overview of the existing data encoding schemes and show how to use them in Genomics. The algorithms provided utilise lossless compression, wavelet-based encoding, and information entropy. Moreover, we propose a contemporary method for testing encoded DNA sequences using Quantum Boltzmann Machines. To evaluate the effectiveness of our algorithms, we discuss a potential dataset that serves as a sandbox environment for testing against real-world scenarios. 
Our research contributes to developing classical-to-quantum data encoding methods in the science of Bioinformatics by introducing innovative algorithms that utilise diverse fields and advanced techniques. Our findings offer insights into the potential of Quantum Computing in Bioinformatics and have implications for future research in this area.
\\
\emph{Keywords}: Data Encoding, DNA Sequencing, Quantum Computing
}
\end{abstract}

\maketitle

\newpage
\listoffigures
\listoftables
\listofalgorithms
\section{Introduction}
\begin{figure}[H]
    \centering
    \includegraphics[scale=0.8]{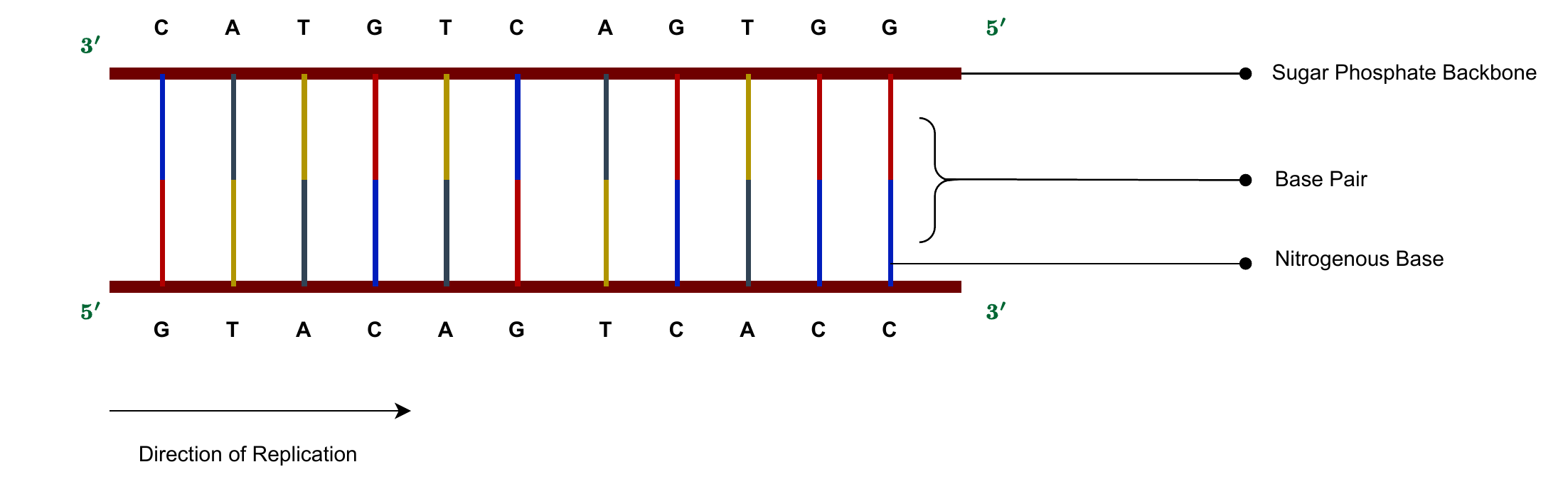}
    \caption{A simplified unravelled DNA sequence.}
    \label{fig:my_label4}
\end{figure}

\textit{Deoxyribonucleic Acid} (DNA) is a molecule that contains the genetic instructions used in the development and functioning of all living organisms. Illustratively, we can describe it as a long, double-stranded molecule made up of nucleotides, which are its composite building blocks. Each nucleotide in DNA consists of the sugar molecule deoxyribose [$\text{C}_{5}\text{H}_{10}\text{O}_{4}$], a phosphate group [--$\text{PO}_{4}^{\;\;3-}$], and a nitrogenous base: Either Adenine [$\text{C}_{5}\text{H}_{5}\text{N}_{5}$], Thymine [$\text{C}_{5}\text{H}_{6}\text{N}_{2}\text{O}_{2}$], Cytosine [$\text{C}_{4}\text{H}_{5}\text{N}_{3}\text{O}$], or Guanine [$\text{C}_{5}\text{H}_{5}\text{N}_{5}\text{O}$], denoted by \textbf{A}, \textbf{T}, \textbf{C}, and \textbf{G} respectively.  The arrangement of these bases along the DNA molecule determines the \textit{genetic code} or \textit{sequence}, which is unique to every individual organism. 
\\
\indent Geometrically, DNA is described as a doubly / bi- helical, which contains two complementary strands of nucleotides that are twisted together as presented in the \textbf{Figure} \ref{fig:my_label4}. The nitrogenous bases of the two strands are held together by Hydrogen bonds, with \textbf{A} always pairing with \textbf{T} and \textbf{C} always pairing with \textbf{G}. This base pairing is often referred to as the base-pairing rule. DNA is found in the nucleus of eukaryotic cells and the cytoplasm of prokaryotic cells. It serves as the template for the synthesis of messenger RNA (mRNA) in a process called transcription, which then serves as the template for protein synthesis in a process called translation. DNA also undergoes replication to produce new copies of itself, ensuring the genetic code is passed on to the next generation of cells and organisms. In addition to its fundamental role in genetics and heredity, DNA has numerous practical applications in Forensics, Medicine, Agriculture, and Biotechnology. \\
\indent The first complete DNA sequence of an organism was determined by the Biochemist Frederick Sanger ($1918-2013$) and his colleagues in 1977. They sequenced the genome of the bacteriophage Phi X 174 ($\phi$X 174) using the Sanger sequencing method, which he and his collaborators had developed a few years earlier \cite{Sanger}. 
\indent Within Genomics, there exist two ubiquitous methods of DNA sequencing (for a complete treatment of the history, see \cite{HistoryOfDNA}): 
\begin{enumerate}
    \item \textbf{Maxam-Gilbert Sequencing:} Also known as ``chemical sequencing'' \cite{MaxamGilbert}, it is a method for determining the sequence of nucleotides in a DNA molecule. This method involves four steps:
    \begin{enumerate}
        \item[1.1.] \textbf{DNA Fragmentation:} The DNA molecule is first broken into smaller fragments of a specific size using chemical or enzymatic methods. 
        \item[1.2.] \textbf{Chemical Treatment:} Each fragment is subjected to a specific chemical treatment that causes DNA cleavage at specific nucleotide positions. There are four chemical reactions, each of which cleaves the DNA at specific nucleotides: Dimethyl Sulfate [$\text{C}_{2}\text{H}_{6}\text{O}_{4}\text{S}$] cleaves at purines \textbf{A} and \textbf{G}, Hydrazine [$\text{N}_{2}\text{H}_{4}$] cleaves at the \textbf{G} residues, Formic Acid [$\text{CH}_{2}\text{O}_{2}$] cleaves at the \textbf{A} residues, and Piperidine [$\text{C}_{5}\text{H}_{11}\text{N}$] cleaves at the \textbf{G} residues.
        \item[1.3.] \textbf{Gel Electrophoresis:} The fragments are then separated by size using polyacrylamide gel electrophoresis (PAGE). The gel is then dried and exposed to X-ray film or a phosphorimager to visualise the labelled fragments.
        \item[1.4.] \textbf{Analysis:} The sequence of the DNA molecule is determined by analysing the pattern of fragments produced by the chemical treatments. By comparing the patterns obtained from each chemical reaction, the location of the cleaved nucleotide can be determined, allowing the sequence of the DNA fragment to be determined.
    \end{enumerate}
    We note that while this was a novel approach to sequencing DNA, it has several drawbacks. For one, it is limited in that it can only effectively sequence short DNA fragments. It is very resource-intensive in its time requirements to carry out this procedure, the hazardousness and toxicity of the chemicals involved, and the amount of material used.  
    \item \textbf{Sanger Sequencing:} Also known as the ``chain-termination'' method \cite{Sanger} or ``Dideoxy Sequencing''. This technique was heavily influenced by, and therefore built upon, the M-G Sequencing technique's bedrock and tried to rectify some of where it falls short. This method finds many applications in Genome Sequencing, Gene Expression Analysis, and Genetic Disease Diagnosis.
    
    Essentially, this technique is ubiquitously used to determine the sequence of nucleotides. The method is based on the incorporation of modified nucleotides called dideoxynucleotides (ddNTPs) into the growing DNA chain during DNA synthesis. These ddNTPs lack a $3'$-Hydroxyl [--OH] group, which is required to form a phosphodiester bond with the next incoming nucleotide. As a result, once a ddNTP is incorporated, no further nucleotides can be added to the DNA chain, effectively terminating the chain elongation.

    Sanger sequencing involves four separate reactions, each containing a different ddNTP and all four regular deoxynucleotides (dNTPs), DNA polymerase, a primer, and the template DNA. These reactions generate a series of DNA fragments of different lengths that terminate at each occurrence of the ddNTP. The fragments are then separated by size using gel electrophoresis, producing a ladder of bands corresponding to the DNA sequence. The sequence can be read by reading the positions of each band in the ladder.

    Modern Sanger sequencing techniques typically use fluorescently-labelled ddNTPs and automated capillary electrophoresis to speed up the sequencing process and increase accuracy.

    Some of the drawbacks of Sanger Sequencing are that it is very expensive to conduct and relatively slow compared to more modern techniques.     
\end{enumerate}
Below we tabulate (\ref{tab:tab1}) some of the modern DNA sequencing methods.
\begin{table}[H] 
\caption{Modern DNA sequencing approaches}
\label{tab:tab1} 
    \centering
    \begin{tabular}{p{5cm}|p{12cm}}
    \hline
    \hline
    \textbf{Technique} &\textbf{Description} \\
    \hline
    Metagenomic Sequencing \cite{Riesenfeld} &This method involves sequencing DNA from environmental samples, such as soil or water, to study microbial communities.  \\
    Single-cell Sequencing \cite{Macosko} &This involves sequencing the DNA of individual cells. This technique finds usage in studying genetic heterogeneity and rare cell populations.  \\
    Next-Generation Sequencing (NGS) \cite{Metzker} &This method includes a range of techniques such as Illumina dye sequencing -- such as Deoxyribonucleoside Triphosphates (dNTP) mix [$\text{C}_{10}\text{H}_{16}\text{N}_{5}\text{O}_{12}\text{P}_{3}$], Ion Torrent sequencing, Pacific Bio (PacBio) sequencing, and Oxford Nanopore sequencing. \\
    Third-Generation Sequencing (TGS) \cite{Li} &This method includes techniques like single-molecule real-time (SMRT) sequencing and nanopore sequencing, which can generate long reads. \\
    Target Sequencing \cite{Krauthammer} &This method involves sequencing specific regions of the genome, such as exomes or panels of genes that are of interest. \\
    Ancient DNA Sequencing \cite{Green} &This method involves sequencing DNA from ancient specimens, such as fossils or mummified remains. \\
    \hline
    \end{tabular}
\end{table}

While the currently existing DNA sequencing techniques have achieved many milestones and have contributed positively to scientific pursuits and enhanced the lives of humanity, these methods suffer from many perils. These include:
\begin{enumerate}
\item \textbf{High Costs:} Despite the rapid reduction in sequencing costs over the past few years, DNA sequencing is still relatively expensive. Several estimates indicate that genome sequencing is generally cheaper than testing a single gene. These studies place the costs in the region of $\$399$-$\$999$ USD.
\item \textbf{Resource Intensive:} DNA samples need to be prepared correctly to ensure accurate sequencing results. For sequencing, the samples that are prepared can be time-consuming and labour-intensive.
\item \textbf{ Incomplete Sequencing:} The process of sequencing DNA can be complex and challenging, and errors may occur at various stages, leading to incomplete or inaccurate sequencing results.
\item \textbf{Errors Arising from Sequencing:} Sequencing techniques may produce errors, leading to inaccuracies in the final sequence data. As an example, consider certain regions of the genome that may be challenging to sequence accurately, which lead to potential errors.
\item \textbf{Requiring Specialist Computing Skills:} The analysis of the large amounts of data generated by DNA sequencing can be challenging and requires specialised expertise with software tools.
\item \textbf{Diversity amongst Individuals when Diagnosing:} There can be significant variations between individuals' DNA sequences, which may make it challenging to identify mutations that cause diseases or other genetic conditions.
\item \textbf{Genome Complexity:} The genome can be complex, and some regions are challenging to sequence due to their repetitive nature or high GC (Guanine-Cytosine) content. This complexity can make it challenging to generate accurate sequence data.
\item \textbf{Sequencing Length Limitations:} Some sequencing methods are limited in the length of DNA fragments that they can sequence, making it challenging to study longer sequences.
\end{enumerate}
\indent Inspired by these pitfalls, we present an ambition to apply Quantum Sciences to try and address these problems. Quantum Information Sciences (QIS), which is the umbrella term that all-encompasses: Quantum Computing (QC), Quantum Error Correction, Quantum Machine Learning, Post-Quantum Cryptography, Quantum Metrology, and Quantum Sensing, has the potential to provide a speedup over some classical computing techniques because it exploits the quantum properties of nature, namely: 
\begin{enumerate}
\item \textbf{Non-locality:} When two or more, particles are intertwined in such a way that measuring the state of one particle expeditiously affects the state of the other particle, irrespective of the distance between the two particles.
\item \textbf{Wave-Particle Duality:} The ability for a particle to exhibit both wave-like and corpuscular properties which manifest based upon how they are observed. 
\item \textbf{Superposition:} The ability for an object, mostly confined to the subatomic scales, to exist in multiple states simultaneously. 
\item \textbf{Entanglement:} More colloquially known as ``spooky action at a distance'', it is the correlation of two or more, particles such that the state of one particle depends on the other, even when these particles may be separated by an infinitely-large distance. 
\item \textbf{Tunneling:} The ability for particles to penetrate through barriers that would be impossible in the realm of Classical Mechanics. 
\item \textbf{Interference:} The phenomenon that occurs when two or more states interact with one another, which leads to the change in probabilities after measurement. 
\item \textbf{Teleportation:} The transmission and conveyance of information from one particle to another without any observable motion of the particles.
\end{enumerate}
\hspace*{0.5cm} Of particular interest in QC is the property of superposition, in which a state $\ket{\psi}$ can be written as a linear combination of the basis states $\ket{0}=\begin{pmatrix} 1 &0 \end{pmatrix}^{T}$ and $\ket{1}=\begin{pmatrix} 0 &1 \end{pmatrix}^{T}$, as $\ket{\psi}=\alpha\ket{0}+\beta\ket{1}$, where $\alpha,\beta\in\mathbb{C}$ are the complex probability amplitudes. Historically, QC has its roots in a $1982$ proposition by the Nobel laureate Richard Feynman ($1918-1988$), who begged the question of whether a computer could harness the properties of quantum mechanics to simulate quantum systems \cite{Feynman}.
\\
\indent In the subsequent years that followed, David Deutsch ($1953-$) and Peter Williston Shor ($1959-$) spearheaded this challenge, and the culmination of the first QC algorithm, ``Shor's algorithm'', was published in $1994$, for the prime factorisation of large numbers. This demonstrated that QC had the potential to provide an exponential speed-up over classical computing in certain tasks; thus, the world took notice. The fundamental unit of QC is the \textit{qubit}, a portmanteau of the words ``quantum'' and ``bit'', which lives in a two-dimensional vector space called the \textit{Bloch sphere}. A qubit is the quantum mechanical analogue of a classical bit, but instead of being in the ``0'' or ``1'' state exclusively, it can simultaneously be in the $0$ and $1$ states, with associated probabilities.  Below, we tabulate (\ref{tab:my_label1}) the various physical implementations of qubits.
\begin{table}[H]
    \caption{List of physical qubit implementations}
    \centering
    \begin{tabular}{p{6cm}|p{6cm}|p{5cm}}
    \hline
    \hline
    \textbf{Physical Implementation} &\textbf{Advantages} &\textbf{Disadvantages} \\
    \hline
    \underline{Trapped Ion Qubits:} Qubits that are formed by trapping ions using a Paul Trap. &Highly stable. &Difficult to upscale. \\
     &Can be precisely manipulated. &Require expensive and complex equipment to implement. \\
     \hline
     \underline{Superconducting Qubits:} Qubits made from superconducting circuits. &Relatively easy to manufacture. &Highly sensitive to noise and decoherence. \\
     &Can be easily upscaled. &Require very low operational temperatures. \\
     \hline 
     \underline{Photon Qubits:} Qubits that use the phase angle or polarisation of individual photons to represent states. &Easy to create. &Difficult to store. \\
     &Easy to manipulate. &Difficult to stimulate interaction with other qubits. \\
     \hline
     \underline{Quantum Dots:} These are minute semiconductors that trap electrons and use their spin to represent qubits. &Easily integrated with current semiconductor technology. &Sensitive to noise and decoherence. \\
     \hline
    \end{tabular}
    \label{tab:my_label1}
\end{table}

\begin{figure}[H]
    \centering
    \includegraphics[scale=0.7]{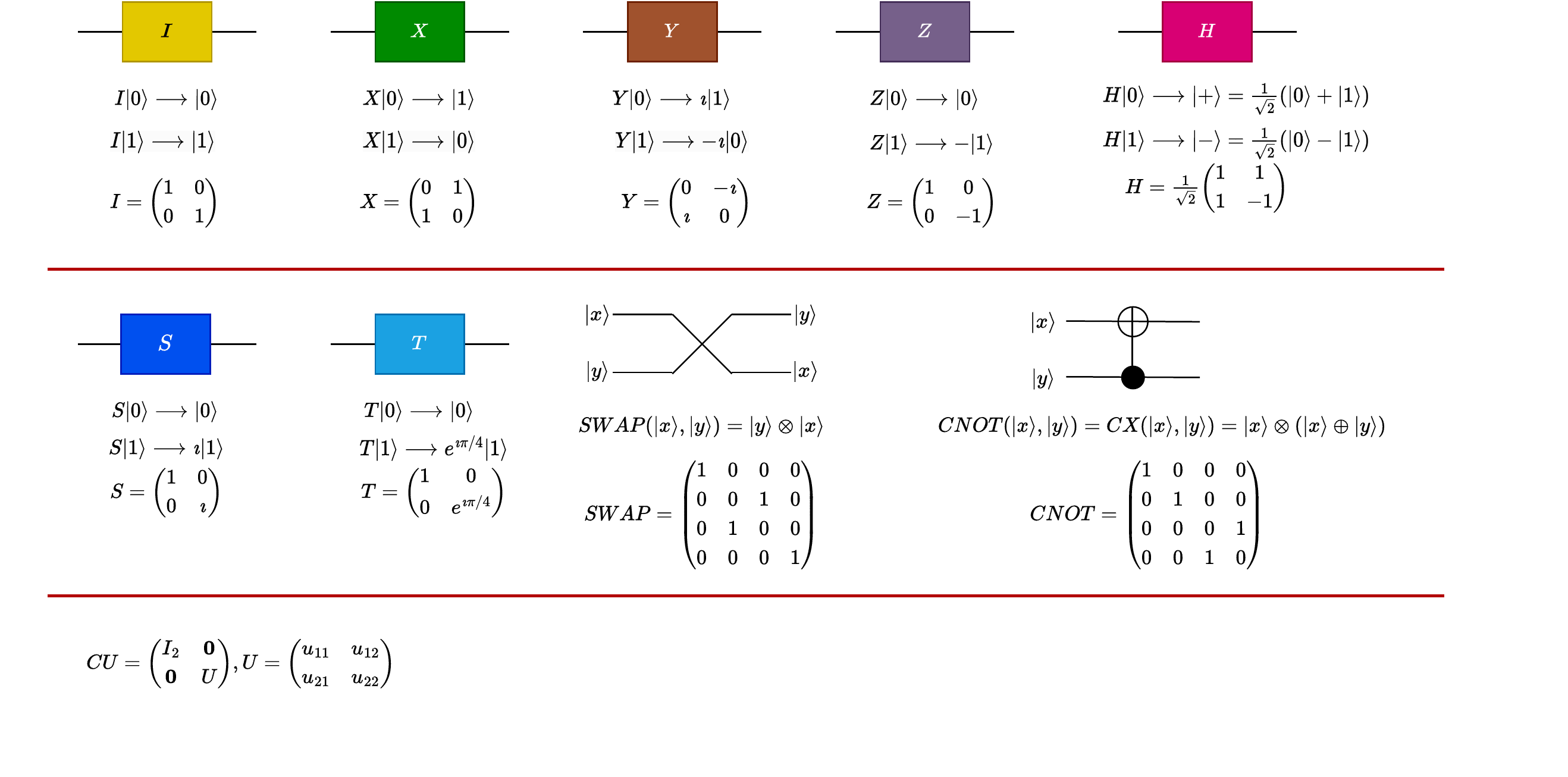}
    \caption{Single and two-qubit quantum gates.}
    \label{fig:my_label7}
\end{figure}

Mathematically, we will restrict our usage and implementation of QC for classical-to-quantum data encoding to gate-based QC; and we note the convention $\imath=\jmath=\sqrt{-1}$; $\imath$ being used in mathematical formulae, and $\jmath$ being used in code output. In addition, we note the in-vogue conventions of $I$ for the identity matrix, $X, Y, Z$ for the  Pauli spin matrices / gates, $H$ for the Hadamard gate, $S$ for the phase gate, $T$ for the $\pi/8$ gate, $SWAP$ for the SWAP gate, $CNOT=CX$ for the controlled-NOT gate, and $CU$ for the controlled-unitary gate, where $U$ is any unitary transformation; \textit{videlicet} $UU^{\dagger}=U^{\dagger}U=I$. \\
\indent Within this archetype, quantum gates, analogous to classical logic gates, serve as representations of matrix operations on states. All classical logic gate results can be represented as well as additional results that arise due to the various quantum phenomena. Below, in the \textbf{Figure} \ref{fig:my_label7}, we diagrammatically encapsulate these gate operations.
Reverting to the discussion on DNA sequencing, to the Physics Community, what was particularly appealing was the determination of the double helix structure of the DNA molecule because one of its co-discoverers, Francis Cricks ($1916-2004$), trained as a Physicist. Nowadays, with the advent of QIS, many Physicists are interested, amongst other research tracks, in using this new computational paradigm for a whole class of novel applications, including, and certainly not limited to, DNA sequencing. 
Within this framework, there has been a plethora of interest in recent times in applying techniques from QC for the sequencing of DNA. \\ 
\indent Kathuria \textit{et al} \cite{Kathuria} presented two efficient inner product-based kernel classifiers for cataloguing individuals as ``normal'' or ``having a disease'' using functional genomic attributes as inputs. The classifiers presented used binary-valued features, which allowed for highly efficient data encoding and Hamming distance-based measurements
\begin{equation}
    d_{H}(s,s')=\sum_{i=1}^{n}\left(s_{i}\neq s'_{i}\right)=\sum_{i=1}^{n}
    \begin{cases} 
        1,\quad s_{i}\neq s'_{i}, \\
        0, \quad\text{otherwise}.
    \end{cases}=\sum_{i=1}^{n}s_{i}\oplus s'_{i},
\end{equation}
where $s_{i}$ and $s'_{i}$ are the two DNA sequences, and $\oplus$ denotes the logical exclusive OR (XOR) operator. Using IBM's quantum computer, the algorithms were implemented. The classifiers required the same number of qubits for training samples and had low gate complexity after the states were prepared. \\
\indent Boev \textit{et al} \cite{Boev} presented a method using quantum and quantum-inspired optimisation techniques to solve genome assembly tasks, with promising experimental results on both simulated data and the $\phi$X 174 bacteriophage. Their results suggested that the new generation of quantum annealing devices could outperform existing techniques for De Novo genome assembly, \textit{videlicet} genome sequencing, where no reference sequences were available to refer to. \\
\indent Sarkar \textit{et al} \cite{Sarkar} presented \textit{QuASeR} as a reference-free DNA sequence reconstruction implementation via De Novo assembly on both gate-based and quantum annealing platforms. Furthermore, four implementation steps with a proof-of-concept example to target the genomics research were presented. \\ 
\indent Na\l\c{e}cz-Charkiewicz and Nowak \cite{NN} presented a proof for a De Novo assembly algorithm using a hybrid combination of CPU and the D-Wave quantum annealer QPU for calculations that were benchmarked against the results of a classical computer by using the Pearson correlation coefficient to detect overlaps between DNA readings. The training data consisted of synthetic and real data from a simulator, and this study was unique because actual organism genomes were sequenced. Due to the low number of qubits available on NISQ-era quantum computers \cite{nisq}, this study demonstrated that hybrid approaches are imperative in the field, and quantum annealers are a viable option for De Novo sequencing. \\
\indent\textit{Data Encoding} or \textit{State preparation}, within the context of QIS, is the process of converting classical data into quantum states for further usage in an algorithm. The method of data encoding occurs within the preprocessing stage. By presenting an efficient encoding scheme, the entire DNA sequencing process will become more streamlined and robust. Thus, enhancing existing quantum sequencing methods or inspiring the development of a rich class of new sequencing algorithms. The domain of this work lies within this process of the quantum data lifecycle. Our foremost objective is to design novel classical-to-quantum encoding schemes such that the data can be used further down the line in the process. 
\begin{figure}[H]
\hspace*{-1cm}
    \centering
    \includegraphics[scale=0.7]{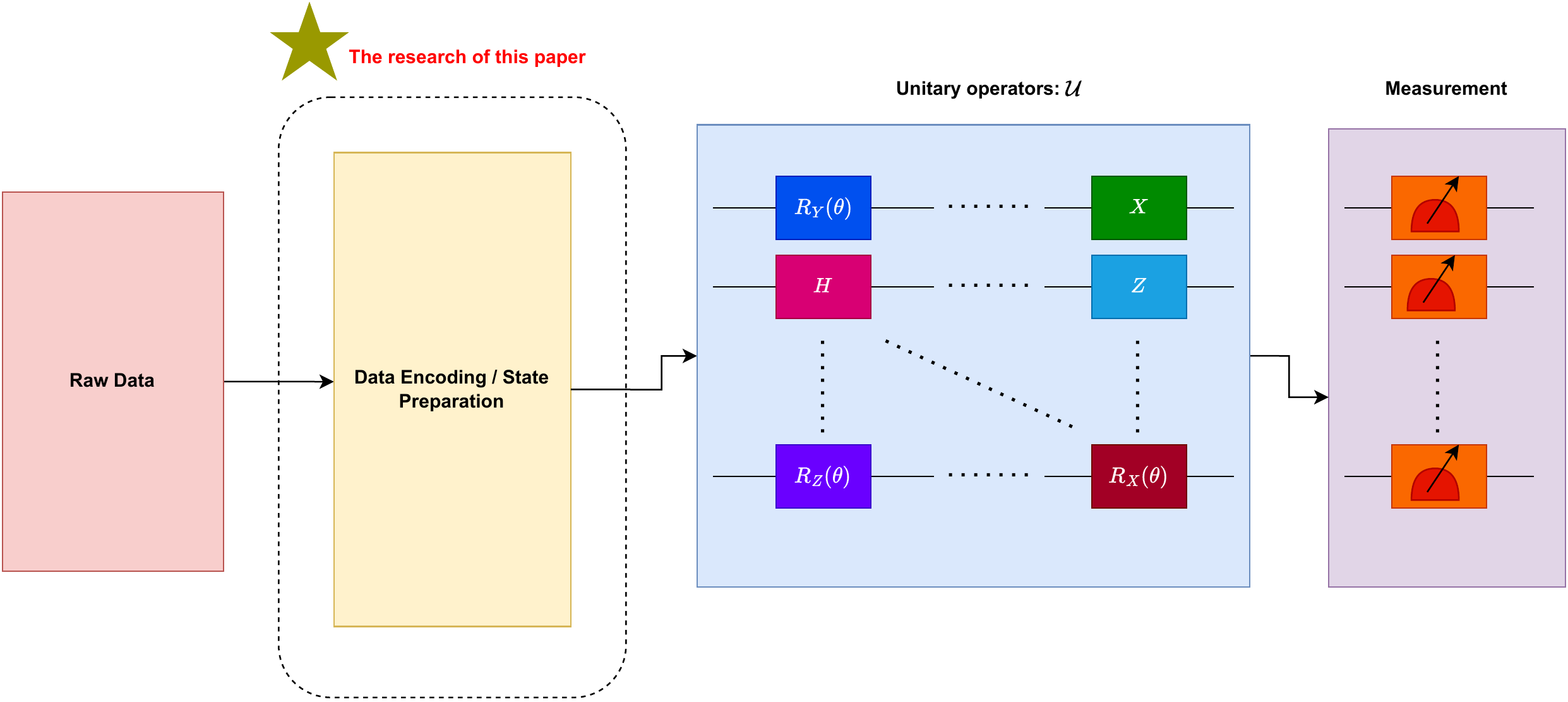}
    \caption{Quantum DevOps lineage of the data lifecycle.}
    \label{fig:my_label5}
\end{figure}
In this paper, we present novel classes of algorithms that are appropriate from methods and ideas in Electrical and Electronic Engineering, Information Theory, Differential Geometry, and Neural Network Architectures. In so doing, we provide seven new algorithms for classical-to-quantum encoding and an eighth algorithm that provides a proof-of-concept test for how effective an algorithm is in undertaking the encoding. 
This paper is divided as follows: \\
 \indent In \S\ref{sec:section1}, we provide an overview of existing classical-to-quantum encoding schemes that are ubiquitously adopted in the discipline, and which are incorporated into several of the popular SDKs and frameworks. \\
\indent In \S\ref{sec:section2}, we provide demonstrable examples of how the existing classical-to-quantum data encoding schemes are used. \\
\indent In \S\ref{sec:section3}, we provide the first classes of new algorithms that use lossless compression. In addition, we motivate why lossless compression-inspired algorithms were chosen over lossy algorithms. \\
\indent In \S\ref{sec:section4}, we provide a wavelet-inspired encoding scheme that, in many respects, builds upon the Quantum Fourier Transform. \\
\indent In \S\ref{sec:section5}, we provide two algorithms based on the notion of information entropy. The input to the first algorithm takes in just the DNA sequence to be encoded, whereas, in the second algorithm, the DNA sequence to be encoded as well as a reference DNA sequence, is ingested. Lastly, using the methods of Computational Information Geometry, we present a third type of procedure. \\
\indent In \S\ref{sec:section6}, we provide a method for testing an encoded DNA sequence using Quantum Boltzmann Machines. \\
\indent In \S\ref{sec:section7}, we discuss a potential dataset that serves as a sandbox environment to test these algorithms against. \\
\indent In \S\ref{sec:section8}, we provide conclusive remarks and contemplate the results produced. Thereafter, an indication of some of the future avenues of research that will be pursued in subsequent iterations of this work are transiently discussed. 
\section{A Review of Existing Encoding Methods}
\label{sec:section1}
Prior to providing an in-depth discussion of the existing techniques, it is imperative that we clearly define ``Rotations''. \textit{Rotation}, in the context of data encoding, is the process of applying a rotation operator / gate, $R_{\xi}(\theta)$, for $\xi\in\left\{X, Y, Z\right\}$, with a phase angle $\theta$ to quantum data. Mathematically, this is achieved as follows:
\begin{equation}
R_{\xi}(\theta)\overset{\Delta}{=}\exp\left(-\imath\frac{\theta}{2}\xi\right)=\cos\left(\frac{\theta}{2}\right)\mathbb{I}_{3}-\imath\sin\left(\frac{\theta}{2}\right)\xi,
\end{equation}
with $\mathbb{I}_{3}$ is the $3\times3$ identity matrix and $\xi$ in this case represents $3\times3$ single Pauli matrix.  
 
Sierra-Sosa \textit{et al.} \cite{sierra2023data} explore rotation schemes on classical data during the preprocessing step of the QML pipeline. It was shown that rotations could have a negative impact on the von Mises dataset, leading to substantial whiskers on the box plots. This outcome is akin to other preprocessing methods, where a chosen normalisation or discretisation can have an unfavourable effect on the classifier's ability to optimise a given dataset.
The existing encoding schemes belong to the following classes: \texttt{Amplitude Encoding} and \texttt{Second-order Pauli Feature Map Encoding}. Below, we discuss each of these classes and their respective components.
\subsection{Amplitude Encoding}
\textit{Amplitude encoding} is a technique used for mapping classical p-dimensional data points $x \in \mathbb{R}^{p}$ to quantum states $\ket{x}$ in the complex Hilbert space $\mathbb{C}^{2n}$ with $n$ is the number of qubits \cite{sierra2023data}. The resulting quantum state is defined by the amplitudes of $x$ in the computational basis states. The original classical data determine these states using the expression below:
\begin{equation}
x \mapsto \ket{x} = \sum_{i=1}^{2^{n}} x_i\ket{i},
\end{equation}
where $\ket{i}$ are the $n$-qubit computational basis states.
We can apply this technique using two classical preprocessing steps that are first carried out. If the input vector $x$ is less than $2^{n}$ in dimension, it is padded with zero features to
create a vector of dimension $2^{n}$. The input vector is normalised to ensure its amplitudes lie in the range $[0,1]$.
\newline
\indent Next, the normalised input vector is transformed into the desired quantum state $\ket{\psi(x)}$ through a series of uniformly controlled rotations on the qubits. The rotation on a given qubit $q_s$ is controlled by all possible states of the previous qubits $q_1,q_{2},\ldots,q_{s-1}$ using multiple controlled rotations. The rotation angle $\beta_{i}$ is determined by the association of the vector $v_i$ representing the $i_{th}$ classical sample with the amplitudes of the original state. The state $\ket{\psi}$ then is defined by this equation:
\begin{equation}
\ket{\psi}=R(v^{i},\beta )\ket{ q_{1}\ldots q_{s-1}}\ket{ q_{s}}.
\end{equation}
Below, we provide the instructions for the 
\texttt{Amplitude Encoding} algorithm:
\begin{algorithm}[H]
\caption{\texttt{Amplitude Encoding}}
\begin{algorithmic}
 \State input a $p$-dimensional classical data point $x\in\mathbb{R}^{p}$, a number $n$ of qubits, and a set of vectors $v_{i}$ representing the classical samples
\State pad the input vector $x$ with zero features as necessary to create a vector of dimension $2^{n}$
\State normalise the input vector $x$ to ensure that its amplitudes lie in the range $[0,1]$
\State initialise the quantum state $\ket{0}^{n}$, where each qubit is
in the state $\ket{0}$
 
\algstore{testcont}
\end{algorithmic}
\label{amplitude}
\end{algorithm}
\begin{algorithm}[H]
\ContinuedFloat
\caption{\texttt{Amplitude Encoding} - Part $2$}
\begin{algorithmic}
\algrestore{testcont} 
\For{each qubit $q_{s}$}
\For{each classical sample $i$}
    \State calculate the rotation angle $\beta_{i}$ based on the amplitudes of the original state and the associated vector $v_{i}$
    \State apply the rotation gate $R(v_i,\beta_{i})$ on qubit $q_s$
  \EndFor
  \EndFor
\State \textbf{return} encoded states $\ket{\psi}$   
\end{algorithmic}
\label{amplitude2}
\end{algorithm} 
Additionally, many studies have shown that QML algorithms exhibit a higher degree of consistency and less variation when the data is encoded using \texttt{Amplitude Encoding} compared to other data encoding methods, such as basis encoding. In particular, the quantum support vector classifier (QSVC) models display a striking absence of variance between data points; see \cite{sierra2023data} for a full elucidation.
\subsection{Encoding via Second-order Pauli Feature Maps}

The Pauli feature map is another technique for encoding classical data into quantum states. We can achieve this method by preparing an initial quantum system and then applying a repeated sequence of unitary operations that involve Hadamard and Pauli gates \cite{sierra2023data}:
\vspace{-0.4\baselineskip}
\begin{equation}
    \mathcal{U}_{\Phi (x)}=U_{\Phi (x)}H^{\otimes n}U_{\Phi (x)}H^{\otimes n},
\end{equation}
where we have that $U_{\Phi (x)}$ can be written as below:
\vspace{-0.4\baselineskip}
\begin{equation}
U_{\Phi (x)}=\exp\left(\imath\sum_{S\subseteq \left [ n \right ] }\phi_{S}(x) \prod_{i\in S}P_{i}\right),
\end{equation}
the Pauli gates $P_{i}$ can be either $I, X, Y,$ or $Z$ gate, and the index $S$ denotes subsets of the set $[n]$ with size $\leqslant k$, describes the degree of connectivity between various qubits or datapoints. The circuit $\phi_{S}$ parameters are defined by a nonlinear function of the input data  as shown in Eq. \eqref{phis} below.
\begin{equation}
\label{phis}
\phi_{S}(x)=
\begin{cases}
x_{0}, \;\quad\quad\quad\quad\quad\quad\text{if}\; k=1, \\
\prod_{j\in S}\left(\pi-x_{j}\right), \quad\text{otherwise}.
\end{cases}
\end{equation}
These parameters separate the different data classes while ensuring an efficient circuit implementation. Below, we delineate \textbf{Algorithm \ref{pauli}}  for the Pauli feature map encoding. 
\begin{algorithm}[H]
\caption{\texttt{Pauli Feature Map Encoding}}
\label{pauli}
\begin{algorithmic}
   \State input: 
    \\ $-$ $n$: number of qubits or data points
    \\ $-$ $x$: input vector of length $n$
    \\ $-$ $k$: maximum size of the subsets $S$
   \State initialise quantum circuit with $n$ qubits
   \State apply a layer of Hadamard gates to all qubits
  \For{each subset $S$ of $\left [ n \right ]$ with size $ \leqslant k$} 
     \State apply a Pauli gate $P_{i}$ to each qubit in $S$, where $P_{i}\in\left\{I,X,Y,Z\right\}$ 
    \State apply a controlled-Z gate between every pair of qubits in $S$.  
    \State apply a nonlinear function of the form $\phi_{S}(x_{i})$ to each qubit in $S$
  \EndFor
  \State apply a layer of Hadamard gates to all qubits
  \State \textbf{return} encoded states $\ket{\psi}$   
\end{algorithmic}
\end{algorithm}
The coefficients $\phi_{S}$ are chosen such that the feature map can efficiently separate the different data classes. For the specific Pauli feature map mentioned in the passage, $k$ is set to $2$, $P_{i}$ is set to $Z$ for each qubit $i$, and $ZZ$ for each pair of qubits $i$ and $j$. The resulting circuit is the \texttt{ZZFeatureMap} in Qiskit and \texttt{IQPEmbedding} in PennyLane.
\subsection{ZZFeatureMap}
\texttt{ZZFeatureMap} is a type of variational quantum circuit in Qiskit that maps classical data to quantum states \cite{ibm}; it applies a series of single-qubit rotations and two-qubit entangling gates to encode the classical data into the quantum state. These gates generate a set of entangled states that can be used as input to a quantum machine learning model.
\newline \indent This feature can be used by defining the number of qubits in the circuit, which is typically determined by the dimensionality of the input data, and then selecting the hyperparameters for the circuit, such as the number of layers, the types of single-qubit rotations, and the types of entangling gates. Finally, we can construct the  \texttt{ZZFeatureMap} circuit. 
\subsection{IQPEmbedding}
\texttt{IQPEmbedding} is a feature in PennyLane that enables the efficient preparation of quantum states for Quantum Phase Estimation (QPE) \cite{pennylane}. The \texttt{IQPEmbedding} circuit takes in a unitary target operator and maps it onto a set of small, highly entangled qubits, allowing for efficient QPE. The \texttt{IQPEmbedding} uses a special type of circuit called an ``interferometer quantum circuit with parameterised gates'' and is commonly used in variational algorithms like the VQE for state preparation. \\
\indent We can use this feature by defining the unitary target operator and selecting the circuit's hyperparameters. Finally, we can construct the \texttt{IQPEmbedding} circuit.
\section{Using the Existing Classical-to-Quantum Encoding Methods on Genomic Data}
\label{sec:section2}
\subsection{Amplitude Encoding}
\begin{figure}[ht] 
\centering
    \includegraphics[scale=0.7]{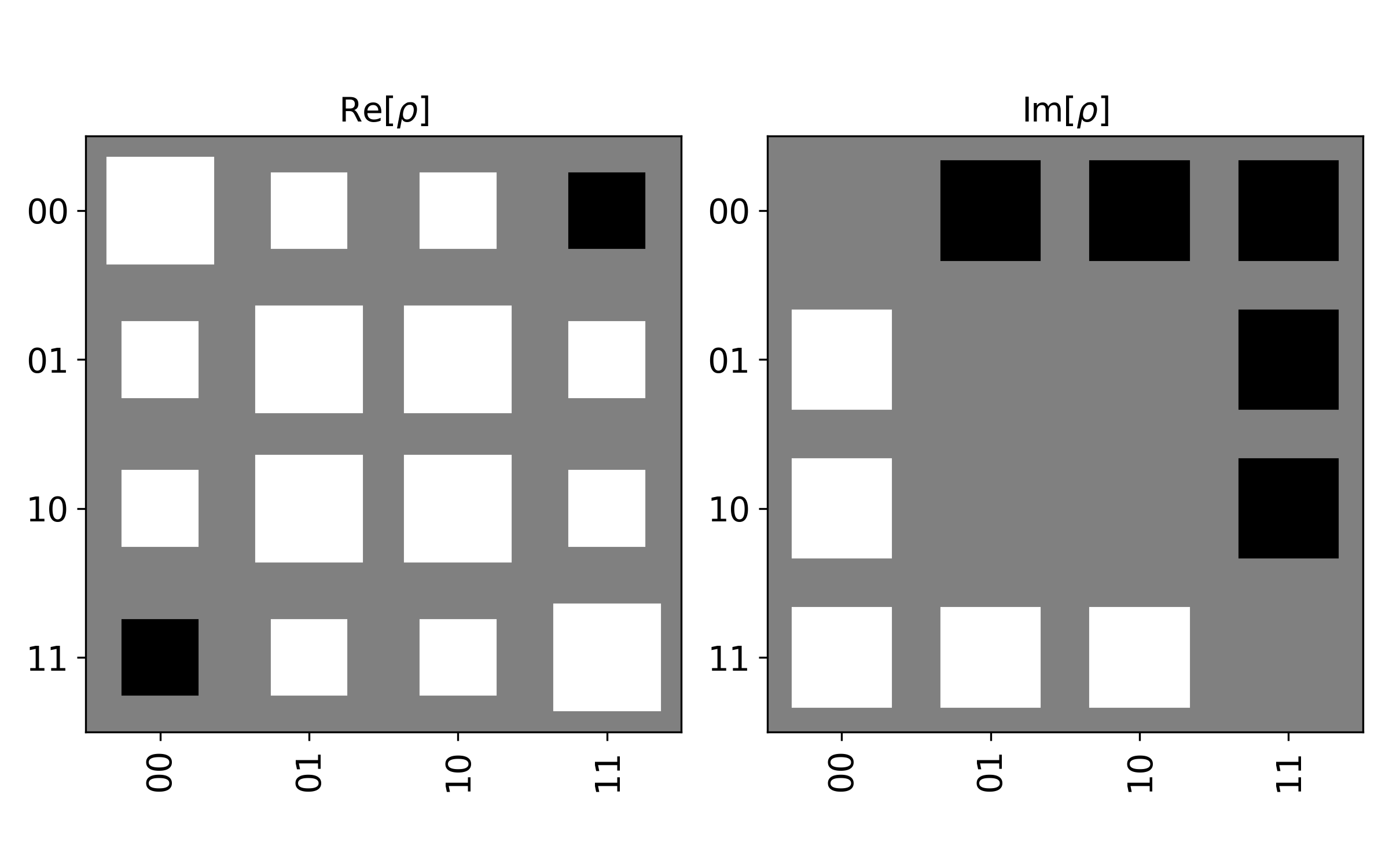}
    \caption{Hinton representation of quantum data for \textbf{A} using \texttt{Amplitude Encoding}, with $\rho$ being the state vector of the sequence \textbf{A} with $\text{Re}(\cdot)$ as the real part and $\text{Im}(\cdot)$ as the imaginary part.}
    \label{fig:amplitudeencoding}
\end{figure}

We implemented the \texttt{Amplitude Encoding} technique using Qiskit to map DNA sequences to quantum states. Specifically, we represented each base in the DNA sequence using a two-bit string and concatenated these strings to form a binary string representing the entire DNA sequence. We then used this binary string as the classical data vector for the \texttt{Amplitude Encoding}. We applied the rotation gates to the qubits and obtained the resulting quantum state vector.
\\
\indent The code is applied in order to map a classical data vector of dimension $2$, representing the DNA sequence \textbf{A}, to a quantum state of $2$ qubits.
In this example, each base in the DNA sequence is represented using two bits: $\textbf{A}=00$, $\textbf{C}=01$, $\textbf{G}=10$, and $\textbf{T}=11$. 
The resulting quantum state is a superposition of all possible combinations of this base, each with a corresponding amplitude; we used the \texttt{statevector()} function from the Qiskit library to represent the resulting quantum state as a vector of complex numbers. 
\\
\indent The output \texttt{statevector} 
\begin{equation*}
\begin{pmatrix}
    0.25-4.33012702\times10^{-01}\jmath & 0.5-7.85046229\times10^{-17}\jmath \\
    0.5-8.32667268\times10^{-17}\jmath & 0.25+4.33012702\times10^{-01}\jmath \\
\end{pmatrix},  
\end{equation*}
shows that the amplitudes of the quantum state are very small, which is expected given the small number of qubits involved; the first element in the first row represents the probability amplitude of the state $\ket{00}$, the second element in the first row represents the state $\ket{01}$, the first element in the second row represents the state $\ket{10}$, and the second element in the second row represents the state $\ket{11}$. This function also provides the dimensions of the quantum state, which is a tensor product of $2$ two-dimensional Hilbert spaces. 
\\
\indent To better visualise this state, we used the Hinton visualisation, which represents the quantum state as a 2D matrix where the size and colour of each square correspond to the magnitude and phase of the amplitude of that state, respectively, as shown in \textbf{Figure} \ref{fig:amplitudeencoding}.
\subsection{Encoding via Second-order Pauli Feature Maps}
In this technique, we represented the genomic data as a binary string, with each nucleotide constituted by a pair of bits. For example, we can represent the sequence as $\textbf{A}=00$, $\textbf{C}=01$, $\textbf{G}=10$, and $\textbf{T}=11$. Then, this binary string was fed into the Pauli feature map circuit, which applied a series of Pauli gates and nonlinear functions to encode the data into a quantum state. The challenge here was the choice of Pauli gate $P_{i}$ and $\phi _{S}$ coefficients and the dyadic choice of two ways to apply the \texttt{ZZFeatureMap} and \texttt{IQPEmbedding}. 
\subsubsection{ZZFeatureMap}
\begin{figure}[ht] 
\hspace*{-6cm}
    \includegraphics[scale=0.7]{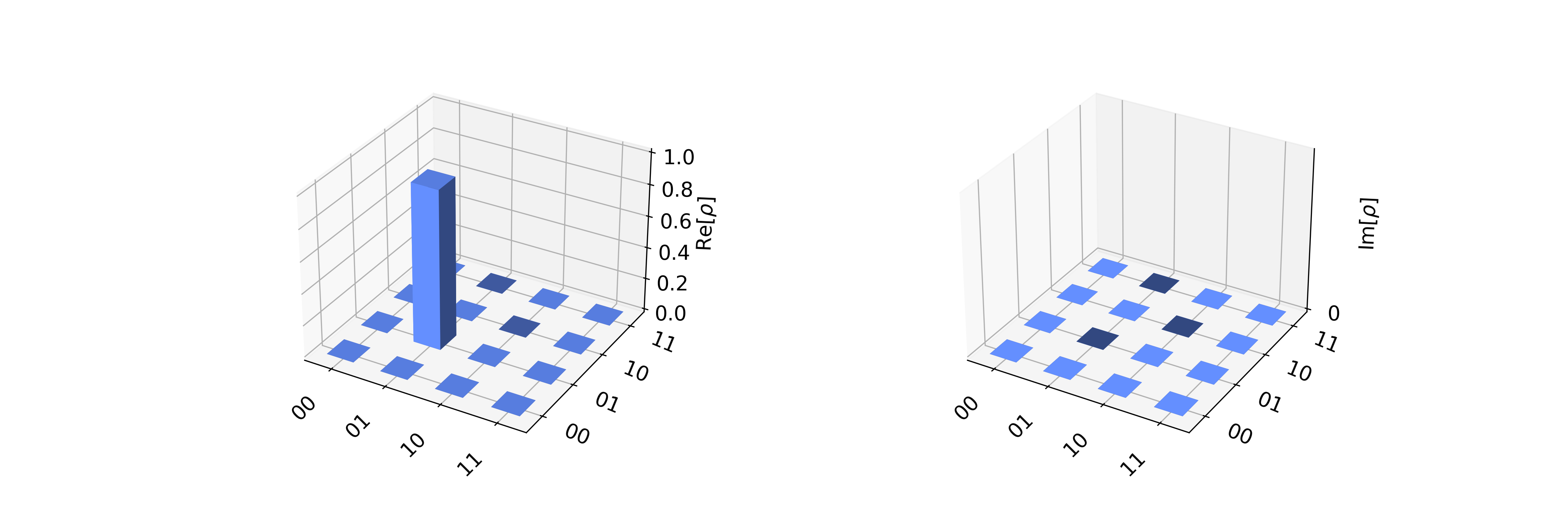}
    \caption{City representation of quantum data for \textbf{AG} using \texttt{ZZFeatureMap}.}
    \label{fig:zzfeaturemap}
\end{figure}

In this technique, the genomic data is converted into a binary representation comparable to the other methods; in our case, the DNA sequence was encoded as a binary string. Then, \texttt{ZZFeatureMap} was used to map this binary string to a quantum state, where each bit in the string corresponds to a single qubit.
\\
\indent 
The code provided \ref{code} demonstrates how to encode a DNA sequence into a quantum state vector using a feature map and a quantum circuit. In this example, the DNA sequence is \textbf{AG}, which implies that the circuit has two qubits, one for each nucleotide.
\\
\indent 
The circuit encodes the DNA sequence by applying gates based on the nucleotides of each qubit. Specifically, gate X is applied to qubits that correspond to nucleotides \textbf{A} or \textbf{C}, and gate $Z$ is applied to qubits that correspond to nucleotides \textbf{G} or \textbf{T}.
\\
\indent 
After applying the \texttt{ZZFeatureMap} to the circuit, as explained in the previous paragraph, the quantum state vector is obtained by executing the circuit on a simulation backend. The resulting state vector is $\left\langle 0.+0.\jmath, 1.+0.\jmath, -0.+0.\jmath, -0.+0.\jmath\right\rangle$, with dimensions $\left (2,2 \right )$.
\\
\indent 
To provide a more precise representation of the quantum state vector, we used the function \texttt{plot\_state\_city} from the \texttt{qiskit$.$visualization} library, In this case, the \textbf{Figure} \ref{fig:zzfeaturemap} shows a dark square on the second row, second column, indicating that the probability of measuring the qubits in the state $\ket{0}$ is $1.0$, which represents the encoded data for the DNA sequence \textbf{AG}.
\subsubsection{IQPEmbedding}
This technique implements a quantum circuit to encode a DNA sequence and obtain the corresponding state vector (\ref{code}). The DNA sequence \textbf{ATG} is taken as input and converted to qubits by the \texttt{AngleEmbedding} circuit. In this circuit, each nucleotide of the DNA sequence is mapped to a qubit position using the \texttt{nucleotide\_map} dictionary. The $R_{Y}$ gate is then applied to each qubit to encode the nucleotide as a rotation angle. 
\\
\indent After encoding the DNA sequence, the $CNOT$ layer is applied to perform entanglement between the qubits. Finally, the state of the qubits is measured, and the state vector is returned. The output is a state vector representing the quantum state of the DNA sequence, with amplitudes for each possible state, as shown in the next expression: 
\begin{equation*}
\begin{pmatrix}
    1.83697020\times10^{-16} + 0\jmath & -2.24963967\times10^{-32} + 0\jmath & -1.22464680\times10^{-16} + 0\jmath \\
    1.00000000 + 0\jmath & 0.00000000 + 0\jmath & 0.00000000 + 0\jmath \\
    0.00000000 + 0\jmath & 0.00000000 + 0\jmath & 0.00000000 + 0\jmath \\
    0.00000000 + 0\jmath & 0.00000000 + 0\jmath & 0.00000000 + 0\jmath \\
\end{pmatrix}. 
\label{matr}
\end{equation*}

\begin{figure}[ht] 
\hspace*{-3cm}
\centering
    \includegraphics[scale=0.7]{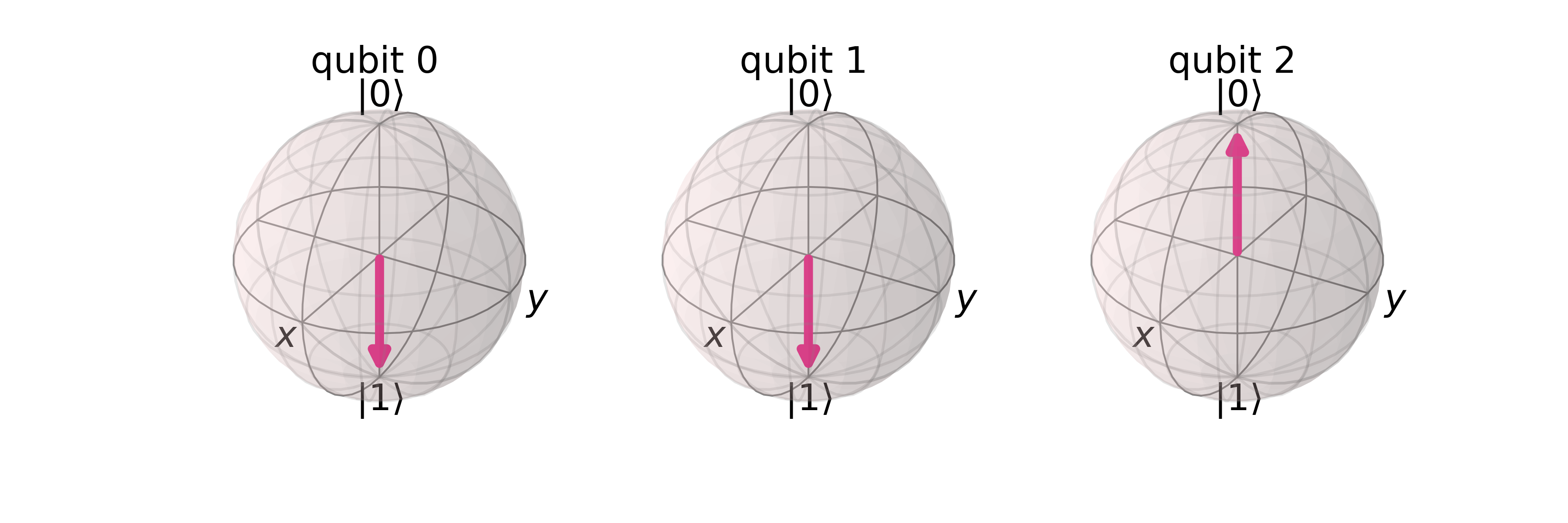}
    \caption{Bloch sphere representation of quantum data for \textbf{ATG} using \texttt{IQPEmbedding}.}
    \label{fig:emb}
\end{figure}

\indent The Bloch sphere representation for each qubit is chosen to understand the results better; the Bloch sphere is used as a visual tool to represent the state of a single qubit and allows for a straightforward interpretation of quantum data, so \textbf{Figure} \ref{fig:emb} represent the Bloch sphere rendition of each qubit in the final \texttt{statevector} and shows that this DNA sequence \textbf{ATG} is converted to $\ket{011}$.
\section{Lossless Compression-Inspired Encoding Schemes}
\label{sec:section3}
In the world of data structures, \textit{compression} minimises the data size to save storage space or transmission time \cite{sayood}. The \textit{modus operandi} of any compression algorithm involves the removal of redundant or irrelevant information from the data without losing quintessential details. This is attained by using various algorithms and techniques that identify patterns or similarities in the data and encode them in a more methodical, and well-ordered, manner.
Data compression algorithms come in two varieties: Lossy and lossless. Below, we provide a curtailed description of both.
\begin{enumerate}
\item \textbf{Lossy Compression:} These algorithms forego some information to achieve a higher compression rate. The compressed data cannot be decompressed to the exact original form, but it is usually acceptable for practical purposes. Examples of lossy compression algorithms include the famous protocols MP3 (MPEG-1/2 Audio Layer III), JPEG, and MPEG.
\item \textbf{Lossless Compression:} Lossless compression algorithms reduce the data size without sacrificing any information, \textit{videlicet} there is no information loss. This implies that the compressed data can be decompressed to the original form without losing quality. Examples of lossless compression protocols include ZIP, GZIP, and PNG.
\end{enumerate}

 Lossless compression algorithms work by identifying patterns in the data and replacing them symbolically with shorter codes, known as \textit{entropy coding}. Based on the aforementioned descriptions, we identify the following, albeit obvious, advantages that lossless compression has over lossy compression and, therefore, resort to developing quantum versions of classical lossless compression algorithms.
\begin{enumerate}
\item \textbf{There is no Information Loss:} The compressed data can be decompressed to its exact original form without any loss of quality. This makes lossless compression suitable for applications where preserving all the information is essential, such as DNA sequencing and encryption, where conserving the original data is indispensable.
\item \textbf{No Perceptual Assumptions are Required:} Lossless compression algorithms do not require assumptions or presuppositions about what parts of the data are more important than others. This makes lossless compression more suitable for data that cannot be easily segmented into perceptually important and unimportant parts, such as DNA sequencing, since it does not make sense to say that one segment of DNA is more ``important'' than the other.
\item \textbf{Recyclability of Compressed Data:} Lossless compression allows for the original data to be reconstructed with no loss of information and, therefore, reused and repurposed. This means that the compressed data can be used repeatedly without any loss in quality, making it suitable for DNA sequencing since a compactified sequence can be used repeatedly.
\item \textbf{Lossless Compression Algorithms Generally Have Lower Computational Complexity as Compared to Lossy Compression Algorithms:} We provide a graphical rendition of some common lossless versus lossy compression algorithms and show how some commonly occurring lossless compression algorithms have a lower computational complexity as compared to the common lossy compression algorithms.
\begin{figure}[t]
    \centering
    \includegraphics[scale=1.0]{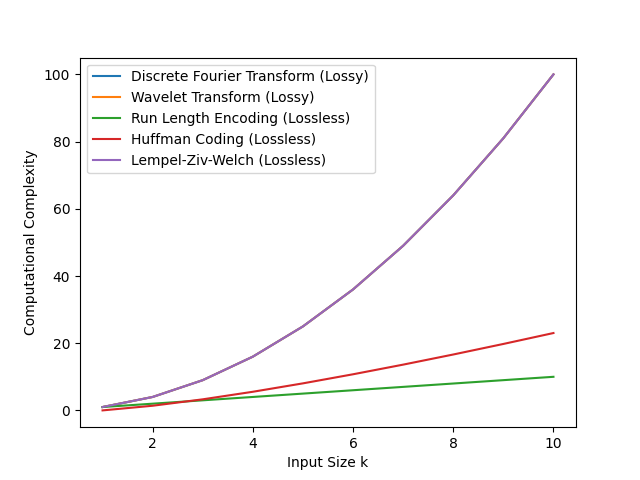}
    \caption{DFT: $\mathcal{O}(k^{2})$; Wavelet Transform: $\mathcal{O}(k^{2})$; Run-length Encoding: $\mathcal{O}(k)$; Huffman Coding: $\mathcal{O}(k\log k)$; LZW: $\mathcal{O}(k^{2})$.}
    \label{fig:my_label1}
\end{figure}
\indent In \textbf{Figure} \ref{fig:my_label1}, we observe that it will be fruitful to design classical-to-quantum encoding schemes based on lossless compression algorithms since the computational time complexities are ordered as follows: $\text{Run-length Encoding}\leq\text{Huffman Encoding}\leq\text{LZW Encoding}=\text{Wavelet Encoding}=\text{DFT Encoding}$. It is important to note that the plot for the DFT Encoding is not displayed due to its resemblance to the graphs of other cases. Consequently, this plot was positioned underneath others and, as a result, was not made visible

\end{enumerate}
\vspace{4cm}
\subsection{A Quantum-influenced Huffman Coding Stratagem}
\textit{Huffman coding} \cite{Huffman} apportions varying-length codes to symbols based on their frequency of occurrence in the input data. At its core, Huffman coding represents frequently occurring symbols in the input data with shorter bit sequences and less frequently repeated symbols with longer bit sequences. In so doing, the algorithm achieves a higher compression ratio than fixed-length coding schemes that use the same number of bits to represent each symbol, regardless of its recurrence.
\\
\indent Architecturally, the algorithm builds a binary tree of nodes, where each leaf node represents a symbol, and its weight is equal to its frequency of occurrence in the input data. The algorithm starts by creating a list of all the symbols and their frequencies and then repeatedly combines the two least frequent symbols into a new node until all the symbols are contained within the tree. The bit sequence for each symbol is then determined by traversing the tree from the root to the corresponding leaf node and assigning a $0$ or $1$ bit for each left or right turn in the path.
In this scheme, we can optimise the number of qubits used in such a manner that the encoded DNA sequence is lean. 
\begin{algorithm}[H]
    \begin{algorithmic}
        \caption{\texttt{QuantHuff}$(\mathcal{D})$ }
\State input a DNA sequence $\mathcal{D}$ 
\For{each nitrogenous base $b$ in the DNA sequence $\mathcal{D}$} 
\State apply \texttt{count}$(b)=C_{b}$ to tally up the frequency of each base 
\State apply \texttt{rank\_order}$(C_{b})=R$ to rank the bases from lowest to highest 
\State  apply \texttt{build\_tree}$(R)=\mathcal{T}$ to construct the root nodes of the Huffman tree 
\State apply \texttt{cascade\_sum}$(\mathcal{T})=S$ to sum up the frequencies in pairs of two from the leftmost side and going up until the apex is reached 
\State apply \texttt{gen\_code}$(S)=G_{b}$ to assign binary digits along the edges, $E$, of the tree 
\If{\text{pos}$(E)$ is left} 
\State $\text{code}\longleftarrow 0$ 
\Else
\State  $\text{code}\longleftarrow 1$ 
\EndIf
\State calculate the number of bits 
\begin{equation*}
    n_{b}=C_{b}\times\;\text{card}(G_{b})
\end{equation*}
\State apply \texttt{bin\_encode}$(G_{b})=\ket{\phi_{b}}$ to convert the bit strings to qubits 
\State calculate the encoded quantum state
\begin{equation*}
\ket{\psi}=\bigotimes\limits_{\text{all bases}}\ket{\phi_{b}}    
\end{equation*}
\EndFor 
\State \textbf{return} encoded states $\ket{\psi}$, total bits required \texttt{sum}$(n_{b})$
\label{QuantHuff}
    \end{algorithmic}
\end{algorithm}
In \textbf{Algorithm \ref{QuantHuff}} above, the code takes in a DNA sequence $\mathcal{D}$ as input, and for each nitrogenous base in the sequence, it counts the frequency of the base, ranks them according to frequency from lowest to highest from left to right. The code then graphically constructs the tree and then sums the frequency in pairs of two starting from the leftmost side, and building its way up until it reaches the tip of the tree. The algorithm then assigns binary bits ($0$ or $1$) to the tree based on the following criteria: If the edge is to the left side, then it is allocated a $0$, and if it is to the right side, it is allocated a $1$. 
\\ 
\indent Thereafter, the number of bits required to encode each nitrogenous base is calculated according to the simple formula of the product between the tally of each base in the sequence with the cardinality of each encoded bitstring of the respective base. Penultimately, the algorithm applies a simple binary encoding function that converts a $0$ into a quantum state $\ket{0}$ and $1$ into a quantum state $\ket{1}$. Lastly, the algorithm calculates the encode quantum state by taking the tensor product of each constituent base state. The algorithm then outputs the encoded quantum state and the sum of the number of bits to give the optimal number of bits required to encode the DNA sequence.
\\
\indent Consider, for example, the M13  filamentous bacteriophage universal reverse primer DNA sequence \cite{m13}: $\boldsymbol{5'-\textbf{CAGGAAACAGCTATGACC}-3'}$ commonly encountered in cloning and PCR testing. Below we apply the encoding scheme, not accounting for the prefix and suffix numbers and hyphens, respectively, and provide a pictorial representation of the algorithm to obtain the encoded quantum state $\ket{\psi}$.

\begin{table}[H]
    \caption{Count of the nitrogenous bases in the M13 bacteriophage DNA sequence}
    \centering
    \begin{tabular}{p{5cm}|p{3cm}}
    \hline
    \hline
    \textbf{Nitrogenous Base} &\textbf{Count} \\
    \hline
     \textbf{C} &5  \\
     \textbf{A} &7 \\
     \textbf{G} &4 \\
     \textbf{T} &2 \\
     \hline
     \textbf{Total} &$\boldsymbol{18}$ \\
     \hline
    \end{tabular} 
    \label{tab:my_label2}
\end{table}
\newpage
\begin{figure}[H]
    \centering
    \includegraphics[scale=1.1]{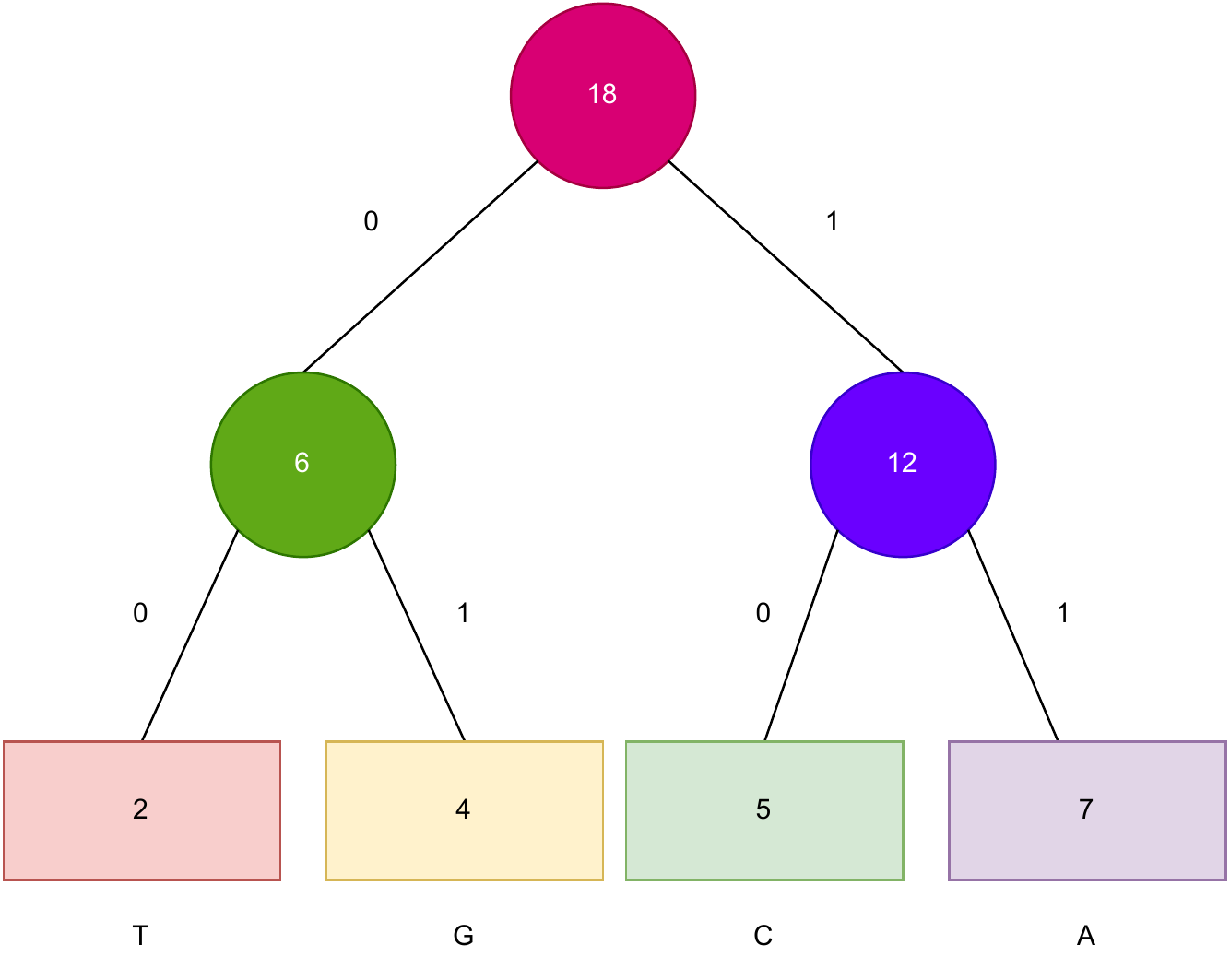}
    \caption{Graphical representation of the nitrogenous bases with their counts and codes along the edges respectively.}
    \label{fig:my_label2}
\end{figure}

\begin{table}[H]
\caption{Code determination for each nitrogenous base.}
    \centering
    \begin{tabular}{p{5cm}p{3cm}p{3cm}p{3cm}}
    \hline
    \hline
    \textbf{Nitrogenous Base} &\textbf{Count} &\textbf{Code} &\textbf{Bits} \\
    \hline
    \textbf{C} &5 &10 &$5\times 2=10$ \\
    \textbf{A} &7 &11 &$7\times 2=15$ \\
    \textbf{G} &4 &01 &$4\times 2=8$ \\
    \textbf{T} &2 &00 &$2\times 2=4$ \\
    \hline 
    \textbf{Total} &$\boldsymbol{18}$ &\textbf{-} &$\boldsymbol{36}$ \\
    \hline 
    \end{tabular}
    \label{tab:my_labe3l}
\end{table}
\vspace{-0.3cm}
Thus, we now apply binary encoding to convert the bits into qubits, \textit{videlicet} $0\longrightarrow\ket{0}, 1\longrightarrow\ket{1}$, to obtain the encoded quantum state for the M13 bacteriophage sequence:
\begin{align*}
\textbf{CAGGAAACAGCTATGACC}\longrightarrow\ket{\psi}=&\;\ket{10}\otimes\ket{11}\otimes\ket{01}^{\otimes 2}\otimes\ket{11}^{\otimes 3}\otimes\ket{10} \nonumber \\
&\;\otimes\ket{11}\otimes\ket{01}\otimes\ket{10}\otimes\ket{00}\otimes\ket{11} \nonumber \\
&\;\otimes\ket{00}\otimes\ket{01}\otimes\ket{11}\otimes\ket{10}^{\otimes 2}.
\end{align*}

\subsection{Quantum-influenced Burrows-Wheeler Transform}
\begin{figure}[ht]

\includegraphics[width=10cm]{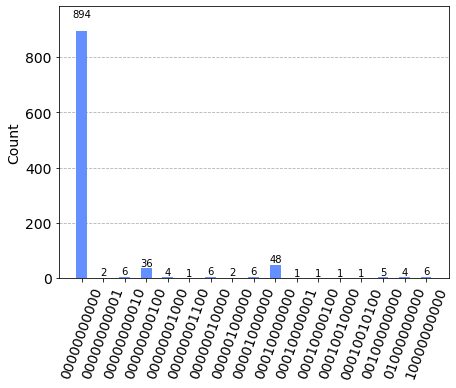}
\caption{The results of \textbf{CTTTTGGAAAA} using \texttt{QBWT} on QasmSimulator with $1024$ shots.}
\label{fig:qbwt}
\centering

\end{figure}

The \textit{Burrows-Wheeler Transform} (BWT) \cite{bw} is a reconstructible text transformation that interchanges the characters in a string in such a manner that it can be compressed  without information loss. The BWT is based on the concept of cyclic shifts, which involve shifting the characters of a string from one position to the left and wrapping the last character to the beginning of the string. In order to successfully carry out a BWT on a string, we first append a special end-of-string marker to the string. Then, a matrix of all the cyclic shifts of the string is constructed and sorted in alphabetical order. The last column of this matrix is the BWT of the string. 
\\
\indent The BWT can be used for compression by exploiting the fact that the last column of the matrix is most often highly repetitive, with runs of identical characters. These runs can be encoded more efficiently than individual characters, resulting in data condensation.
Within the Biosciences, the BWT is in the construction of the Burrows-Wheeler Aligner (BWA) software invented by Li and Durbin in the $\texttt{C++}$ language \cite{LiDurbin}, which is an exhaustively used tool for mapping DNA sequences to a reference genome. Owing to the fact that DNA sequences contain large amounts of repetitive nitrogenous bases and, therefore, repetitious information, the BWT can be used to exploit this redundancy and compress the sequence without losing information. 
\\
\indent This compression can reduce storage requirements and improve the speed of sequence analysis. In addition, the BWT is used in De Novo genome assembly \cite{KimuraKoike}, Sequence Annotation \cite{Adjeroh}, and Motif Finding (repeating patterns in DNA) \cite{motif} for the construction of data structures. 
Taking motivation from its immense success in various DNA sequencing applications, we combine quantum computing in parallel with the BWT protocol for encoding.  Below, we present a novel classical-to-quantum encoding scheme that uses the BWT.
\newpage
\begin{algorithm}[H]
\caption{\texttt{QBWT}$(\mathcal{D})$}
\begin{algorithmic}
\State input a DNA sequence $\mathcal{D}\in\mathbb{S}$
\State initialise a quantum state $\ket{\psi}=\ket{0}^{\otimes n}$
\For {each nitrogenous base in $\mathcal{D}$}
\State apply \texttt{cyc\_perm}$(\mathcal{D})=\mathcal{D}'$ exhaustively, until there are no repeats
\State apply \texttt{arrange}$(\mathcal{D}')=\mathcal{D}''$ to alphabetically arrange $\mathcal{D}'$
\State apply \texttt{extract}$(\mathcal{D}'')=\mathcal{D}'''$ to obtain the last nitrogenous bases, and index $j$ of the original DNA sequence from $\mathcal{D}''$
\State apply the \texttt{concat}$(\mathcal{D}''',j)$ to link together the terminal bases and index of the original DNA sequence
\For{each nitrogenous base  $b\in\mathcal{D}'''$}
\State apply the unitary transformation 
\begin{equation*}
U_{b}=\exp\left[\frac{2\pi\imath\times\texttt{count}(b)}{n}\right]\times R_{b}   
\end{equation*}
\EndFor
\EndFor
\State \textbf{return} encoded states $\ket{\psi}$
\end{algorithmic}
\label{agbwt}
\end{algorithm}
\indent In \textbf{Algorithm \ref{agbwt}}, we observe that the algorithm possesses quadratic time complexity, $\mathcal{O}(k^{2})$. Upon analysis of \textbf{Algorithm \ref{agbwt}}, we observe that the first for loop is the steps for the classical BW transform, and the second for loop is the quantum part. We also take heed of the following notations: $\mathbb{S}$ denotes the space of all possible DNA sequences, $n$ is the number of qubits, the function \texttt{cyc\_perm}$(\cdot)$ computes the number of cyclic permutations of the DNA sequence, the function \texttt{arrange}$(\cdot)$ provides an alphabetically-ordered arrangement of all the permutations of the DNA sequence, the \texttt{extract}$(\cdot)$ function draws out all the terminal nitrogenous bases from the alphabetically-ordered permutations, the function \texttt{count}$(b)$ counts the number of occurrences of nitrogenous base $b$ in $D'''$ and $R_{b}$ is the rotation operator that maps $\ket{0}\longrightarrow\ket{b}$.
\\
\indent The code \ref{code} provided implements the \texttt{QBWT} algorithm using the Qiskit library for quantum computing \cite{ibm}. The \texttt{QBWT} function takes a DNA sequence as input and performs the classical BWT step using the arrange function to generate the transformed sequence. It then initialises a quantum circuit with $n$ qubits (where $n$ is the length of the sequence) and applies a series of quantum rotations to each qubit based on the frequency of each base in the transformed sequence. Finally, it measures all the qubits and returns the counts of each measurement outcome.
\\  
\indent This implementation of the \texttt{QBWT} algorithm provides a simple example of how we can encode DNA sequences into quantum data. 
The results of the \texttt{QBWT} algorithm are the counts of the measurement outcomes of the final quantum state after the quantum phase estimation step. These counts represent the probabilities of observing each possible outcome when measuring the state and can be interpreted as an encoded representation of the input DNA sequence. The counts represent the frequency of each combination of bases in the input DNA sequence.
\\
\indent For example, if the input DNA sequence is \textbf{ACTGACGTAGC}, it has a length of $10$, so the quantum circuit has ten qubits. The circuit first applies the Burrows-Wheeler Transform (BWT) classically to the input sequence, which produces a new string where each character corresponds to the last character of one of the cyclic permutations of the input sequence, arranged in alphabetical order.
\\
\indent In this case, the BWT of \textbf{ACTGACGTAGC} is \textbf{CTTTTGGAAAA} since the cyclic permutations are \textbf{ACTGACGTAGC}, \textbf{CTAGCAGTACG}, \textbf{GACGTAGCACT}, \textbf{TAGCACTGACG}, \textbf{AGCACTGACGT}, \textbf{GTAGCACTGAC}, \textbf{ACGTAGCACTG}, \textbf{CACTGACGTA}, \textbf{TTTGGAAAAA}, and \textbf{AAAACTTTTG}, and the last character of each permutation, arranged alphabetically, is \textbf{CTTTTGGAAAA}.
\\
\indent The circuit then applies the quantum phase estimation algorithm to the qubits, where each qubit corresponds to one base in the input sequence. For each base in the input sequence, the algorithm counts the number of occurrences of that base in the BWT and uses that count to apply a rotation gate to the corresponding qubit. The rotation angle is proportional to the count of the base divided by the length of the input sequence and is given by the formula: $\theta = 2 * \pi * count / n$, where $n$ is the length of the input sequence.
\\
\indent Finally, the circuit measures all of the qubits and returns the counts of each measurement outcome, where each count corresponds to a binary string of length $n$. 
In this case, as shown in \textbf{Figure} \ref{fig:qbwt}, we have $18$ different binary strings, each representing a different state of the quantum circuit. The most common state, measured $894$ times, is the all-zero state $0000000000$, which corresponds to the case where none of the qubits is rotated. The least common states, each measured only once, are $0000100100$, $0001010010$, $0011000000$, and $0100000000$.

\section{Wavelet-Inspired Encoding Schemes}
\label{sec:section4}
\textit{Wavelets} are mathematical functions that can be used to analyse and transform signals in both the time and frequency domains. Ubiquitously, they are encountered in the field of signal processing due to their ability to provide more detailed information about a signal than traditional Fourier transforms. The most common type of wavelet is the \textit{mother wavelet}, which can be scaled and translated in order to analyse signals at different frequencies and time intervals. The wavelet transform involves taking the inner product of the signal with the scaled and translated wavelet function. Mathematically, a wavelet can be expressed in terms of its transform, known as the \textit{wavelet transform}. For a given mother wavelet $\Psi(t)$ and signal $f(t)$, the wavelet transform is given by
\begin{equation}
W(a;b)\overset{\Delta}{=}\frac{1}{\sqrt{|a|}}\int_{-\infty}^{+\infty}f(t)\Psi^{*}\left(\frac{t-b}{a}\right)dt,
\end{equation}
where $a$ is the scale factor, $b$ is the translation factor, and the $^{*}$ operator denotes the complex conjugate. Whereas a Fourier transform is stationary in time, and therefore can only account for localised signal variations, wavelet transforms are dynamic in time, and thus can encapsulate both localised, as well as globalised, changes in the signal. In addition, a wavelet can simultaneously capture temporal and frequency information, whereas a Fourier transform can only provide information about frequency. \\
\indent Below, we present a wavelet-inspired encoding scheme that can handle DNA sequence image data; in stark contrast to the other schemes presented which attempt to encode the DNA sequences directly. This scheme can work well with image data, processing, compression, and analysis.
For an $M\times N$-dimensional grayscale image data, we provide the \textbf{Algorithm \ref{algi}}  for encoding the classical data into quantum states:

\begin{algorithm}[H]
\caption{\texttt{CosineEncoding}$(\mathcal{D})$ }
\begin{algorithmic}
\State input an $\left(M\times N\right)$-dimensional grayscale image dataset $\mathcal{D}\in\mathbb{R}^{M\times N}$
\For {each row index in $M$, and each column index in $N$}
\State calculate the two-dimensional Discrete Cosine Transform of the image according to
\begin{align*}
F(\alpha,\beta)=&\;\frac{C_{\alpha}}{2}\frac{C_{\beta}}{2}\sum_{x=0}^{M-1}\sum_{y=0}^{N-1}f(x,y)\cos\left[\frac{\alpha\pi\left(2x+1\right)}{2M}\right]\cos\left[\frac{\beta\pi\left(2y+1\right)}{2N}\right], \\
C_{\alpha}, C_{\beta}=&\;
\begin{cases}
\frac{1}{\sqrt{2}}, \quad \alpha,\beta=0, \\
1, \;\;\quad otherwise.
\end{cases}
\end{align*}
\State find $F_{\max}=\underset{\alpha\in M,\beta\in N}{\max} ||F(\alpha,\beta)||$
\State normalise each value according to $\widehat{F}(\alpha,\beta)=\frac{F(\alpha,\beta)}{F_{\max}}$
\For{each element in $\widehat{F}{i}(\alpha,\beta)$}
\State map $||\widehat{F}(\alpha,\beta)||\longrightarrow$  $a_{i}, 1\leq i\leq\text{card}(\widehat{F}(\alpha,\beta))=n$ to form an $n$-qubit register
\State apply a QFT to each state in the n-qubit register:
\begin{equation*}
    \text{QFT}_{n}\ket{p}=\frac{1}{\sqrt{n}}\sum_{q=0}^{n-1}\exp\left(\frac{2\pi\imath pq}{n}\right)\ket{q}.
\end{equation*}
\EndFor
\EndFor
\State \textbf{return} encoded $2^{n}$-dimensional dataset of quantum states $\mathcal{D}'\in\mathcal{H}$
\end{algorithmic}
\label{algi}
\end{algorithm}

\begin{figure}[H] 
\centering
    \includegraphics[scale=0.7]{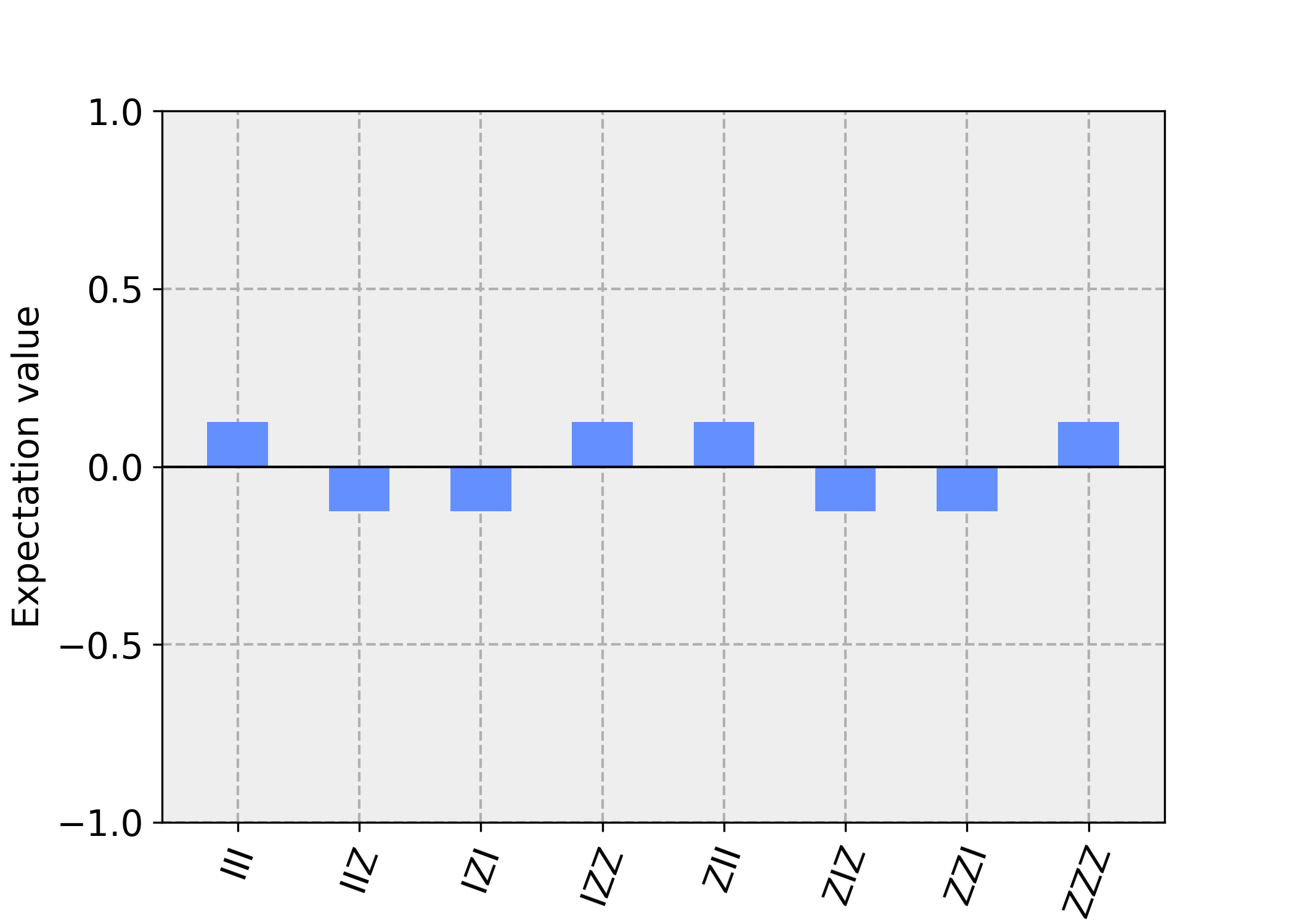}
    \caption{Pauli representation of quantum data for \textbf{ATC} using \texttt{Cosine Encoding}.}
    \label{fig:cosine}
\end{figure}

\begin{figure}[ht] 
\hspace*{-6cm}
    \includegraphics[scale=0.7]{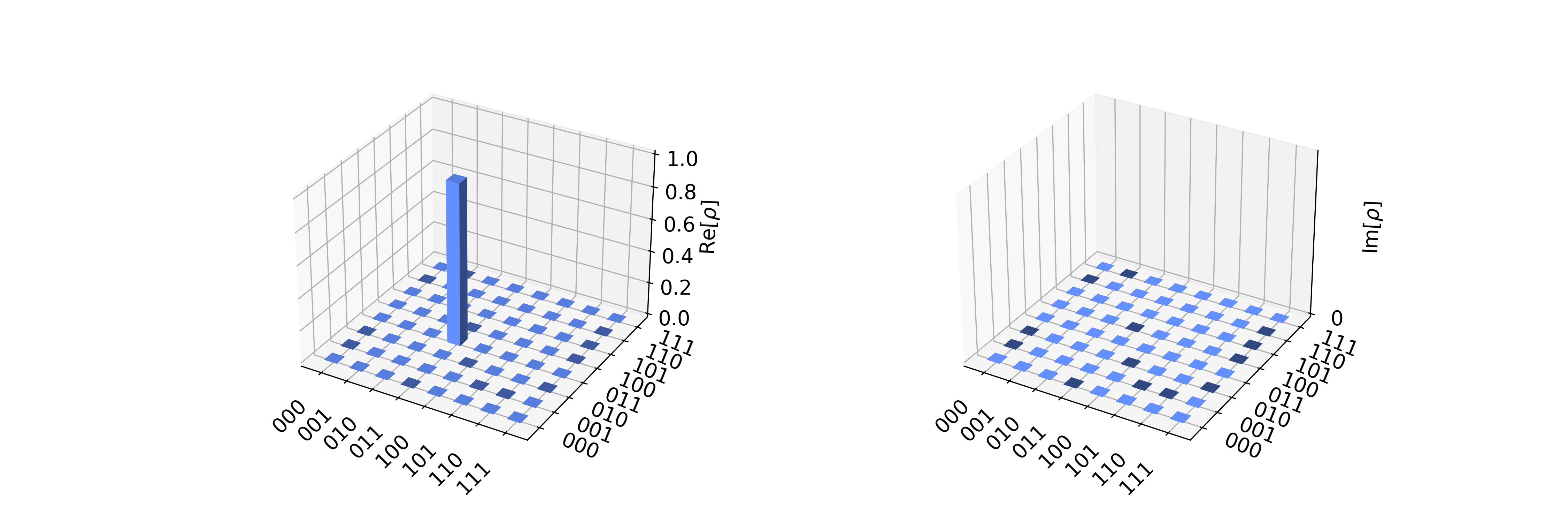}
    \caption{City representation of quantum data for \textbf{ATC} using \texttt{Cosine Encoding}.}
    \label{fig:cosine2}
\end{figure}

In \textbf{Algorithm \ref{algi}}, $f(x,y)\in\left[0,255\right]$ denotes the intensity of the pixels for an 8-bit value, $\alpha,\beta$ are the positions, and $C_{\alpha}, C_{\beta}$ are the scale factors. We also observe that this algorithm displays quadratic time complexity, $\mathcal{O}(k^{2})$. 
This algorithm encodes a DNA sequence into a quantum state using a quantum circuit. The input DNA sequence is converted into a quantum state by initialising each base as a quantum state. If the base is \textbf{A} or \textbf{G}, it is initialised as the $\ket{0}$ state, while if it is \textbf{C} or  \textbf{T}, it is initialised as the $\ket{1}$ state. For \textbf{G} and \textbf{T}, an $X$ gate is applied before initialisation to convert the $\ket{1}$ state to the $\ket{0}$ state.
\\
\indent
The quantum Fourier transform is then applied to each qubit of the circuit, transforming the quantum state into a superposition of all possible states. This is achieved by applying a series of Hadamard and controlled-phase gates to each qubit of the circuit.
Finally, the qubits are measured to obtain classical data, and the \texttt{statevector} of the circuit is obtained using the \texttt{StatevectorSimulator}  from the Qiskit Aer package. The \texttt{statevector} represents the probability amplitudes of each possible system state.
\\
\indent
For example, consider the DNA sequence \textbf{ATC}. The quantum state obtained after encoding this sequence is given as the \texttt{statevector} 
\begin{equation*}
\left\langle 0.+0.\jmath, -0.+0.\jmath, 0.+0.\jmath, 0.70710678+0.70710678\jmath, -0.+0.\jmath, 0.+0.\jmath, 0.+0.\jmath, 0.-0.\jmath\right\rangle, 
\end{equation*}
this implies that a $70.7\%$ probability of the third qubit being in the state $\ket{1}$ and a $50\%$ probability of the first qubit being in the state $\ket{0}$. 
\\
\indent 
The \texttt{plot\_state\_paulivec} function was used to visualise the quantum state as a vector of probabilities along the Bloch sphere's $I$, $X$, $Y$, and $Z$ axes, as shown in \textbf{Figure} \ref{fig:cosine}.
By representing the state vector of a multi-qubit system in the Pauli basis, we can see how much of each Pauli matrix is present in the state.
So in our example, the state vector corresponds to the DNA sequence \textbf{ATC}, which has a length of $3$. The resulting \texttt{statevector} has eight components since there are three qubits, each of which can be in either the $\ket{0}$ or $\ket{1}$ state. Plotting the state vector in the Pauli basis shows that the state mainly comprises the $Z$ matrix, with some contribution from the $X$ matrix; this tells us that the state is mainly in the $\ket{0}$ state for all three qubits, with a small probability of being in the $\ket{1}$ state for the second qubit, as we can see this in the \textbf{Figure} \ref{fig:cosine2}.
\section{Entropy-based Methods}
\label{sec:section5}
In information theory, \textit{entropy} is defined as the average amount of information contained in each message or symbol in a message. This average is calculated based on the probability distribution of the symbols in the message. The entropy of a system is the greatest when all possible messages are equally likely, and least when a message is certain to occur. Within ML, entropy is used to measure the impurity of a dataset. The information gain is calculated as the difference between the original dataset's entropy and the predicted values output from the model. ML algorithms based upon entropy methods try to maximise the information gain, leading to a more pure (less uncertain) dataset in the resulting groups. \\
\indent Entropy-based methods for encoding are a natural consequence when comparing an encoded DNA sequence to a reference DNA sequence. It is also reasonable to assume that DNA sequences have a high degree of inherent randomness. \\
\indent Thus, we provide a class of algorithms built upon entropy-based schemes below.
\vspace{-\baselineskip}
\subsection{Encoding Based on the Lowest Entropy, and Fixed DNA subsequence Lengths}    
\begin{algorithm}[H]
\caption{\texttt{SEncode}$(\mathcal{D}, n)$}
\begin{algorithmic}
\State input a DNA sequence $\mathcal{D}\in\mathbb{S}$ of length $N, |\mathcal{D}|=N$ and the number of qubits $n$
\State initialise an $n$-qubit register in the $\ket{0}$ state, $\ket{0}^{\otimes n}$
\For {each nitrogenous base $b_{i}\in\mathcal{D}$}
\State calculate the probability of occurrence of the bases
\begin{equation*}
    p(b_{i})=\frac{\text{Number of occurrences of}\;b_{i}}{N}, \quad b_{i}\in\left\{\textbf{A},\textbf{T},\textbf{C},\textbf{G}\right\}
\end{equation*}
\State calculate the information entropy of the entire DNA sequence
\begin{equation*}
    H(\mathcal{D})=-\sum_{i\in\mathcal{D}}p(b_{i})\log_{2}\left[p(b_{i})\right]
\end{equation*}
\State calculate $K=\lceil \log_{2}(N)\rceil, M=\lceil\frac{N}{K}\rceil$
\State apply \texttt{fragment}$(\mathcal{D})$ to subdivide the DNA sequence into $M$ segments each of size $K$ such that $\mathcal{D}=\bigcup_{i=1}^{M}\mathcal{D}_{i}$
\For {each fragmented DNA sequence}
\State calculate the probability of the nitrogenous base in the segment
\State calculate the entropy of each segment
\State calculate the maximum entropy from all segments
\begin{equation*}
    H_{\max}=\max\left\{H(\mathcal{D}_{1}), H(\mathcal{D}_{2}),\ldots, H(\mathcal{D}_{M})\right\}
\end{equation*}
\State normalise entropy in each segment
\begin{equation*}
    \widehat{H}(\mathcal{D}_{i})=\frac{H(\mathcal{D}_{i})}{H_{\max}}
\end{equation*}
\State apply the \texttt{sort\_low\_high}$(\widehat{H}(\mathcal{D}_{i}))=\mathscr{H}$ to order the normalised entropies from lowest to highest
\algstore{testcont}
\end{algorithmic}
\label{SEncode}
\end{algorithm}
\begin{algorithm}[H]
\ContinuedFloat
\caption{\texttt{SEncode}$(\mathcal{D}, n)$ - Part $2$}
\begin{algorithmic}
\algrestore{testcont} 
\For{each segmented DNA sequence $\mathcal{D}_{i}$ corresponding to $\mathscr{H}$}
\State synthesise quantum states $\ket{\psi_{i}}$ corresponding to each DNA segment
\State apply a unitary operator to map $\ket{0}\longrightarrow \ket{i}$ 
\begin{equation*}
    U_{i}:\ket{0}\longrightarrow\ket{i}\otimes\ket{0}
\end{equation*}
\State apply a Hadamard gate to each qubit in the first register to create a superposition
\EndFor
\State form the encoded state
\begin{equation*}
    \ket{\psi}=\bigotimes\limits_{i=1}^{M}\ket{\psi_{i}}
\end{equation*}
\EndFor
\EndFor
\State \textbf{return} encoded states $\ket{\psi}$ 
\end{algorithmic}
\label{SEncode2}
\end{algorithm}
\textbf{Algorithm \ref{SEncode}}  takes a DNA sequence $\mathcal{D}$ of length $N$ and the number of qubits $n$ as input and produces an encoded quantum state $\ket{\psi}$ as output. The encoding is done by mapping each segment of the DNA sequence into a corresponding quantum state.
The \texttt{SEncode} algorithm subdivides the DNA sequence into $M$ segments, each of size $K$. The algorithm then calculates the probability of occurrence of each nitrogenous base in the DNA sequence, the entire DNA sequence's information entropy, and each segment's entropy. The maximum entropy among all the segments is then used to normalise the entropy in each segment. The normalised entropies are then sorted from lowest to highest.
\\
\indent The algorithm synthesises a corresponding quantum state for each segment and applies a unitary operator to map $\ket{0}$ to the corresponding segment index. A Hadamard gate is applied to each qubit in the first register to create a superposition. Finally, the algorithm forms the encoded quantum state by taking the tensor product of all the synthesised quantum states.
\\
\indent For example, consider the DNA sequence $\mathcal{D}=$ \textbf{ATCG}, the number of nitrogenous bases $N=4$, we assume that we want to encode the DNA sequence into a $2$-qubit quantum state. The algorithm would first subdivide the DNA sequence into two segments of size $2$: $\mathcal{D}_{1}=$ \textbf{AT} and $\mathcal{D}_{2}=$ \textbf{CG}. The algorithm would then calculate the probability of occurrence of each nitrogenous base, the information entropy of the entire DNA sequence, and the entropy of each segment. 
\\
\indent As an example, assume that the normalised entropy of $\mathcal{D}_{1}$ is $0.5$, and the normalised entropy of $\mathcal{D}_{2}$ is $0.5$. The algorithm would synthesise quantum states corresponding to each segment by applying a unitary operator to map $\ket{0}$ to the corresponding segment index and applying a Hadamard gate to each qubit in the first register to create a superposition; we get the quantum states $\ket{\psi_1}=\frac{1}{\sqrt{2}}\left(\ket{0}+\ket{1}\right)$ and $\ket{\psi_2}=\frac{1}{\sqrt{2}}\left(\ket{0}-\ket{1}\right)$, corresponding to the segments $\mathcal{D}_1=$ \textbf{AT} and $\mathcal{D}_2=$ \textbf{CG}, respectively. Finally, the algorithm would form the encoded quantum state as $\ket{\psi}=\ket{\psi_1}\otimes\ket{\psi_2}=\frac{1}{2}\left(\ket{00}+\ket{01}-\ket{10}-\ket{11}\right)$. 
\subsection{Encoding Based on Reference Data}
\begin{algorithm}[H]
\begin{algorithmic}
\caption{\texttt{NZ22}$(\mathcal{D}, \mathcal{D}_{\text{ref}})$ }
\State input a DNA sequence $\mathcal{D}$ and a reference DNA sequence $\mathcal{D}_{\text{ref}}$ of the same length, $|\mathcal{D}|=|\mathcal{D}_{\text{ref}}|=N$ 
\For{each nitrogenous base $b_{i}\in\mathcal{D}, b_{i}^{\text{ref}}\in\mathcal{D}_{\text{ref}}$ }
\State calculate the probability of occurrence of the bases
\begin{align*}
    P(b_{i})=&\;\frac{\text{Number of occurrences of}\;b_{i}}{N}, \quad b_{i}\in\left\{\textbf{A},\textbf{T},\textbf{C},\textbf{G}\right\} \\
    Q(b_{i}^{\text{ref}})=&\;\frac{\text{Number of occurrences of}\;b_{i}^{\text{ref}}}{N}, \quad b_{i}^{\text{ref}}\in\left\{\textbf{A},\textbf{T},\textbf{C},\textbf{G}\right\} 
\end{align*}
\State calculate the KL divergence between the $\mathcal{D}$ and $\mathcal{D}_{\text{ref}}$ probability distributions
\begin{equation*}
D_{KL}(P||Q)=\sum_{\text{all bases}}P(b)\log\left[\frac{P(b)}{Q(b)}\right]
\end{equation*}
\State calculate the number of qubits required for some adjustable hyperparameter $0\leq\alpha\leq 1$
\begin{equation*}
n=\alpha\times\lceil D_{KL}(P||Q)\rceil 
\end{equation*}
\State create the quantum state by mapping $\ket{\psi}:\mathcal{D}\longrightarrow\mathcal{D}_{\text{ref}}$ in $n$-qubit registers 
\EndFor 
\State \textbf{return} encoded quantum state $\ket{\psi}$     
\label{NZ22}
\end{algorithmic}

\end{algorithm}

While this \textbf{Algorithm \ref{NZ22}} may be effective, we identify that the choice of the KL metric has several flaws. We discuss them below:
\begin{enumerate}
\item It does not account for overlaps between the DNA and DNA reference sequences probability distributions.  
\item It is not symmetrical. This means that the order of the respective probability distributions matter and, thus, significantly impact the results.
\item Since we are mapping the DNA sequence to the computational basis $\left\{\ket{0}, \ket{1}\right\}$, it makes sense to have a divergence measure that is in the range $\left[0,1\right]$. However, the KL metric is $\left[0, \infty\right)$.
\end{enumerate}

Let us consider a practical example to understand how this algorithm works; suppose we have two DNA sequences, $\mathcal{D}$ and $\mathcal{D}{\text{ref}}$, of length $N=1000$ nucleotides each. We want to encode $\mathcal{D}$ into a quantum state using $\mathcal{D}{\text{ref}}$ as a reference sequence. We apply the \texttt{NZ22} algorithm as follows:
\\
\indent First, we calculate the probability of occurrence of each nitrogenous base in $\mathcal{D}$ and $\mathcal{D}_{\text{ref}}$. For example, if we have $300$ Adenine (\textbf{A}) nucleotides in $\mathcal{D}$, the probability of occurrence of \textbf{A} in $\mathcal{D}$ is $P(\textbf{A}) = \frac{300}{1000} = 0.3$. Similarly, if we have $250$ Thymine (\textbf{T}) nucleotides in $\mathcal{D}_{\text{ref}}$, the probability of occurrence of \textbf{T} in $\mathcal{D}_{\text{ref}}$ is $Q(\textbf{T}) = \frac{250}{1000} = 0.25$. We calculate these probabilities for all nitrogenous bases in both sequences.
\\
\indent Next, we calculate the KL divergence between the probability distributions of $\mathcal{D}$ and $\mathcal{D}_{\text{ref}}$. The KL divergence measures how different two probability distributions are from each other. In our example, suppose we calculate $D_{KL}(P||Q) = 0.05$, which indicates that the probability distributions of $\mathcal{D}$ and $\mathcal{D}_{\text{ref}}$ are somewhat different from each other.
\\
\indent We then calculate the number of qubits required to encode $\mathcal{D}$ into a quantum state. We use an adjustable hyperparameter $0\leq\alpha\leq 1$ to control the number of qubits. In our example, suppose we set $\alpha = 1$. We calculate the number of qubits required as $n = \alpha \times \lceil D_{KL}(P||Q) \rceil = 1 \times \lceil 0.05 \rceil = 1$. This means we need one qubit to encode $\mathcal{D}$ into a quantum state.
\\
\indent Finally, we create the quantum state by mapping the nucleotides of $\mathcal{D}$ to the reference sequence $\mathcal{D}_{\text{ref}}$ using the one qubit register. The mapping is done in such a way that if a nucleotide in $\mathcal{D}$ matches a nucleotide in $\mathcal{D}_{\text{ref}}$, the qubit state is set to $\ket{0}$, and if they don't match, the qubit state is set to $\ket{1}$. This creates an encoded quantum state $\ket{\psi}$ that represents the differences between $\mathcal{D}$ and $\mathcal{D}_{\text{ref}}$.
We propose \textbf{Algorithm \ref{NZ23}}  that uses the Bhattacharyya divergence measure \cite{Bhatt1, Bhatt2} to overcome the drawbacks of the KL divergence alluded to above.
\begin{algorithm}[H]
\caption{\texttt{NZ23}$(\mathcal{D},\mathcal{D}_{\text{ref}})$}
\begin{algorithmic}
\State Input a DNA sequence $\mathcal{D}$ and a reference DNA sequence $\mathcal{D}_{\text{ref}}$ of the same length, $|\mathcal{D}|=|\mathcal{D}_{\text{ref}}|=N$
\For {each nitrogenous base $b_{i}\in\mathcal{D}, b_{i}^{\text{ref}}\in\mathcal{D}_{\text{ref}}$}
\State Calculate the probability of occurrence of the bases
\begin{align*}
    P(b_{i})=&\;\frac{\text{Number of occurrences of}\;b_{i}}{N}, \quad b_{i}\in\left\{\textbf{A},\textbf{T},\textbf{C},\textbf{G}\right\} \\
    Q(b_{i}^{\text{ref}})=&\;\frac{\text{Number of occurrences of}\;b_{i}^{\text{ref}}}{N}, \quad b_{i}^{\text{ref}}\in\left\{\textbf{A},\textbf{T},\textbf{C},\textbf{G}\right\} 
\end{align*}
\State Calculate the Bhattacharyya divergence between the $\mathcal{D}$ and $\mathcal{D}_{\text{ref}}$ probability distributions
\begin{equation*}
    D_{B}(P||Q)=-\log\left[\sum_{\text{all bases}}\sqrt{P(b)\times Q(b)}\right]
\end{equation*}
\State Calculate the number of qubits required for some adjustable hyperparameter $0\leq \alpha\leq 1$ 
\begin{equation*}
n=\alpha\times\lceil D_B(P||Q)\rceil
\end{equation*}
\State Apply \texttt{Amplitude Encoding} to obtain state $\ket{\psi}$ based on $n$ qubits
\EndFor
\State \textbf{return} encoded states $\ket{\psi}$ 
\end{algorithmic}
\label{NZ23}
\end{algorithm}
Suppose we have a DNA sequence $\mathcal{D}$ of length $N=10$: $\mathcal{D} = \textbf{TACAGTTGCA}$, We also have a reference DNA sequence $\mathcal{D}_{\text{ref}}$ of the same length: $\mathcal{D}_{\text{ref}} = \textbf{AGCTGACTCA}$

To encode this DNA sequence into a quantum state using \texttt{NZ23}, we follow the steps outlined in \textbf{Algorithm \ref{NZ23}}.

Firstly, we calculate the probability distributions of each base in the input sequence and the reference sequence. For example, the probability of the base \textbf{T} occurring in $\mathcal{D}$ is:
\begin{equation*}
P(\textbf{T}) = \frac{\text{number of \textbf{T}'s in }\mathcal{D}}{N} = \frac{3}{10} = 0.3.
\end{equation*}
Similarly, we can calculate the probability distribution of each base in the reference sequence $Q(b_{i}^{\text{ref}})$.

Next, we calculate the Bhattacharyya divergence between the probability distributions of $\mathcal{D}$ and $\mathcal{D}_{\text{ref}}$. This tells us how different the two probability distributions are, which gives us an estimate of how many qubits we need to encode the DNA sequence into a quantum state.

For this example, the Bhattacharyya divergence is:
\begin{equation*}
D_{B}(P||Q) = -\log \sum_{b\;\in\;\left\{\textbf{A}, \textbf{C}, \textbf{G}, \textbf{T}\right\}} \sqrt{P(b) \times Q(b)} \approx 0.01.
\end{equation*}

The next step is determining the qubits needed to encode the DNA sequence. This is determined by a hyperparameter $\alpha$ between 0 and 1 and can be adjusted to trade-off between accuracy and computational resources. For this example, let's set $\alpha = 1$.

Using the Bhattacharyya divergence and $\alpha$, we calculate the number of qubits required as:
\begin{equation*}
n = \alpha \times \lceil D_B(P||Q) \rceil = 1 \times \lceil 0.01 \rceil = 1.
\end{equation*}
This means we need one qubit to encode this DNA sequence into a quantum state.

Finally, we obtain the quantum state using \texttt{Amplitude Encoding}. For this example, we can encode the state $\ket{\psi}$ as:
\begin{equation*}    
\ket{\psi} = \sqrt{P(\textbf{T})} \ket{\textbf{T}} + \sqrt{1 - P(\textbf{T})} \ket{\neg\textbf{T}} = \sqrt{0.3} \ket{\textbf{T}} + \sqrt{0.7} \ket{\neg\textbf{T}},
\end{equation*}
where $\ket{\neg\textbf{T}}$ denotes a superposition of the other three bases (\textbf{A}, \textbf{C}, and \textbf{G}), and $\neg$ is the logical NOT operator.
\\
\indent So, the \texttt{NZ23} algorithm takes a DNA sequence and a reference sequence as input, calculates the probability distributions of each base, computes the Bhattacharyya divergence between the two probability distributions, and uses amplitude encoding to encode the DNA sequence into a quantum state. The number of qubits required for encoding is determined by a hyperparameter $\alpha$.
\subsection{Information Geometry Methods}
Computational information geometry is the amalgamation of Differential Geometry and Information Theory that studies the properties of probability distributions, generalised entropy measures, and their relationships to Statistics, Probability Theory, Information, Machine Learning, Deep Learning, and Big Data \cite{Frank}. Given a manifold $\mathscr{M}\left(X,g\right)\in C^{\gamma}$ that is smooth and potentially infinitely differentiable, we would like to explicate an information metric, $g_{ij}$, to define the concept of \textit{distance} between probability distributions; a common information distance measure is the Fischer Information Metric: For a continuous random variable $X$ with parameterised probability distribution $p_{\boldsymbol{\theta}}(x)$:
\vspace{-0.3\baselineskip}
\begin{equation}
g_{ij}(\boldsymbol{\theta})=\int_{X}p_{\boldsymbol{\theta}}(x)\nabla_{\boldsymbol{\theta}_{i}}p_{\boldsymbol{\theta}}(x)\nabla_{\boldsymbol{\theta}_{j}}p_{\boldsymbol{\theta}}(x)\;dx.
\end{equation}
Thereafter, an entropy measure is defined to measure the dissimilarity and discrepancy between probability distributions; most commonly, this is the Kullback-Liebler or Jensen-Shannon divergences: Given two continuous probability distributions $p(x)$ and q(x), 
\begin{align}
D_{KL}(p(x)||q(x))=&\;\int_{\mathscr{M}}p(x)\log\left[\frac{p(x)}{q(x)}\right]\;dx, \tag{10.1}\label{10.1} \\
D_{JS}(p(x)||q(x))=&\;\frac{D_{KL}(p(x)||m(x))+D_{KL}(q(x)||m(x))}{2}, \tag{10.2}\label{10.2} 
\end{align}
where $m(x)=\frac{p(x)+q(x)}{2}$. Below, we provide the \textbf{Algorithm \ref{quantig}} for classical-to-quantum encoding using information geometry.
\begin{algorithm}[H]
\caption{\texttt{QuantIG}$(\mathcal{D}, \mathcal{D}_{\text{ref}})$}
    \begin{algorithmic}
 \State input a DNA sequence $\mathcal{D}$ and a reference DNA sequence $\mathcal{D}_{\text{ref}}$ of the same length $|\mathcal{D}|=|\mathcal{D}_{\text{ref}}|=N$
\For{each base $b_{i}\in\mathcal{D}, b_{i}^{\text{ref}}\in\mathcal{D}_{\text{ref}}$}
\State calculate the probability of occurrence of the bases
\begin{align*}
p(b_{i})=&\;\frac{\text{Number of occurrences of}\;b_{i}}{N}, \quad b_{i}\in\left\{\textbf{A},\textbf{T},\textbf{C},\textbf{G}\right\} \\
q(b_{i}^{\text{ref}})=&\;\frac{\text{Number of occurrences of}\;b_{i}^{\text{ref}}}{N}, \quad b_{i}^{\text{ref}}\in\left\{\textbf{A},\textbf{T},\textbf{C},\textbf{G}\right\}
\end{align*}
\State calculate an information metric, $g$, between the probability distributions $p$ and $q$
\State construct a Riemannian manifold $\mathscr{M}(X,g)$ for the probability distributions
\State perform state mapping from the manifold $\mathscr{M}$ to a Hilbert space $\mathcal{H}$ that creates a basis: $\ket{x}:\mathscr{M}\longrightarrow\mathcal{H}$
\State calculate the linear operator
\begin{equation*}
\mathscr{L}\ket{x}=\sqrt{p(x)}\ket{x}
\end{equation*}
\State calculate the metric operator
\begin{equation*}
\mathscr{G}\ket{x}=g(x,x)\ket{x}\bra{x}+\sum_{x\neq y}g(x,y)\ket{x}\bra{y}
\end{equation*} 
\State calculate the quantum states
\begin{equation*}    \ket{\psi}=\mathscr{L}^{-1/2}\mathscr{G}\mathscr{L}^{-1/2}\ket{0}
\end{equation*}
  \EndFor
\State \textbf{return} encoded states $\ket{\psi}$   
    \end{algorithmic}
\label{quantig}
\end{algorithm}

\indent The \texttt{QuantIG} algorithm takes a DNA sequence and a reference sequence as inputs, calculates the probability of occurrence of each base and constructs a Riemannian manifold for the probability distributions. It then maps the state from the manifold to Hilbert space and calculates the quantum states using linear and metric operators. The resulting encoded states can be used for various quantum information processing tasks like similarity search and classification. The manifold can be constructed as follows:
\begin{enumerate}
    \item \textbf{Defining the Probability Space:} Pick a non-varying sample space $\Omega$, or a subset $\omega\subseteq\Omega$, that is bounded and smooth, over the probability distributions.
    \item \textbf{Defining the Tangent Space:} Consider small perturbations of the probability distributions while preserving the normalisation constants. 
    \item \textbf{Metric Construction:} This is accomplished by defining a suitable distance measurement, depending on the application, that satisfies the properties of being smooth and continuous, positive-definite, and having a symmetric bilinear form, for the probability distributions. Given the probability distributions $P$ and $Q$, with associated tangent vectors $X$ at tangent space $T_{P}$, and $Y$ at tangent space $T_{Q}$ respectively, and metric $g$, the commonly used metrics are:
    \begin{enumerate}
        \item[3.1.] \textbf{Fischer-Rao Metric:} $g_{P}(X,Y)\overset{\Delta}{=}\frac{1}{4}\int\frac{d\log P}{d x}\frac{d\log Q}{dx}\;dx$.
        \item[3.2.] \textbf{Wasserstein / Earth Mover's Distance / Optimal Transport:} Measures the amount of ``work'' required to transform one probability distribution into the other: $g_{\mathcal{P}}(P,Q)\overset{\Delta}{=}\inf\left\{\int||x-y||^{\mathcal{P}}\;d\gamma(x,y)\right\}^{1/\mathcal{P}}$, where $\mathcal{P}$ is a control parameter that regulates the smoothness of the distance, and $\gamma(x,y)$ is a joint probability distribution function on the product space $x\times y$ that has $P$ and $Q$ are marginal probability distributions.
        \item[3.3.] \textbf{Hellinger Distance:} $g(P,Q)\overset{\Delta}{=}\frac{1}{\sqrt{2}}\int\left[\sqrt{P(x)}-\sqrt{Q(x)}\right]^{2}\;dx$.
    \end{enumerate}
\end{enumerate}
We note the caveat that the process described above is for continuous probability distributions. In DNA sequencing, discrete probability distributions are encountered. Thus, we approximate a continuous manifold as a discrete manifold using piecewise linear approximations. 
\\ 
\indent For example, consider a small DNA sequence of length $10$: \textbf{ATCGATCGAT}. Let the reference sequence be \textbf{ATCGTTAGCT}. Using the \textbf{Algorithm \ref{quantig}}, we first calculate the probability distributions for the sequence and the reference:
\begin{align*}
p(\textbf{A})=&\;0.3, \quad\quad q(\textbf{A})=0.4, \\
p(\textbf{T})=&\;0.4, \quad\quad q(\textbf{T})=0.2, \\
p(\textbf{C})=&\;0.2, \quad\quad q(\textbf{C})=0.2, \\
p(\textbf{G})=&\;0.1, \quad\quad q(\textbf{G})=0.2.
\end{align*}
\indent Using these probabilities, we construct a discrete manifold for the distributions and perform state mapping to obtain a basis in a Hilbert space. We then calculate the linear and metric operators, then use their values to get our encoded states. This is achieved as follows: The discrete probability distributions are
\begin{align*}
P=&\;\left\{p(\textbf{A}),p(\textbf{T}),p(\textbf{C}),p(\textbf{G})\right\}=\left\{0.3,0.4,0.2,0.1\right\}, \\
Q=&\;\left\{q(\textbf{A}),q(\textbf{T}),q(\textbf{C}),q(\textbf{G})\right\}=\left\{0.4,0.2,0.2,0.2\right\}.
\end{align*}
Associated with the probability distributions, we define the manifold $\mathcal{M}$ with tangent vectors
\begin{equation*}
X=\left(dx,dy,dz,dw\right), \quad\quad Y=\left(dx',dy',dz',dw'\right),
\end{equation*}
where $d\varphi$ are infinitesimal perturbations for $\varphi\in\left\{x,x',y,y',z,z',w,w'\right\}$. We define a potential tangent space for each tangent vector as
\begin{align*}
T_{P}=&\;\left\{\left(dx,dy,dz,dw\right)|dx+dy+dz+dw=0\right\}, \\
T_{Q}=&\;\left\{\left(dx',dy',dz',dw'\right)|dx'+dy'+dz'+dw'=0\right\}.
\end{align*}
Lastly, we compute the Fischer-Rao metric as follows:
\begin{align*}
g_{P}(X,Y)=&\;\sum_{i\in\left\{\textbf{A},\textbf{T},\textbf{C},\textbf{G}\right\}}\frac{1}{p(i)}\times\frac{dp(i)}{dx}\times\frac{dq(i)}{dx} \\
=&\;\sum_{i,j,k\in\left\{\textbf{A},\textbf{T},\textbf{C},\textbf{G}\right\}}\frac{1}{p(i)}\left\{\delta_{ij}dx_{i}dx_{j}+\frac{1}{p(j)}\left[\delta_{ij}dx_{i}dx_{j}-p(i)\right]\right\}\left\{\delta_{ik}dx_{i}dx_{k}+\frac{1}{q(k)}\left[\delta_{ik}dx_{i}dx_{k}-q(i)\right]\right\},
\end{align*} 
where $\delta_{ij}, \delta_{ik}$ are Kronecker delta functions. Simplifying and substituting the probability values, we obtain
\begin{align*}
g_{P}(X,Y)=&\;\left(\frac{1}{0.3}\right)\left(\frac{1}{0.4}\right)dx^{2}+\left(\frac{1}{0.4}\right)\left(\frac{1}{0.2}\right)dy^{2}+\left(\frac{1}{0.2}\right)\left(\frac{1}{0.2}\right)dz^{2}+\left(\frac{1}{0.1}\right)\left(\frac{1}{0.2}\right)dw^{2} \\
&\;+\frac{2\left[\left(0.3\right)\left(0.4\right)+\left(0.2\right)\left(0.1\right)-2\left(0.3\right)\left(0.2\right)\right]}{\left(0.3\right)\left(0.4\right)\left(0.2\right)\left(0.2\right)}dydz \\
=&\;\frac{25}{3}\;dx^{2}+\frac{25}{2}\;dy^{2}+25\;dz^{2}+50\;dw^{2}+\frac{25}{3}\;dydz.
\end{align*}


\section{Energy-Based Methods for Testing Encoded Sequences}
\label{sec:section6}
\begin{figure}[ht]
\hspace*{-1cm}
    \centering
    \includegraphics[scale=0.6]{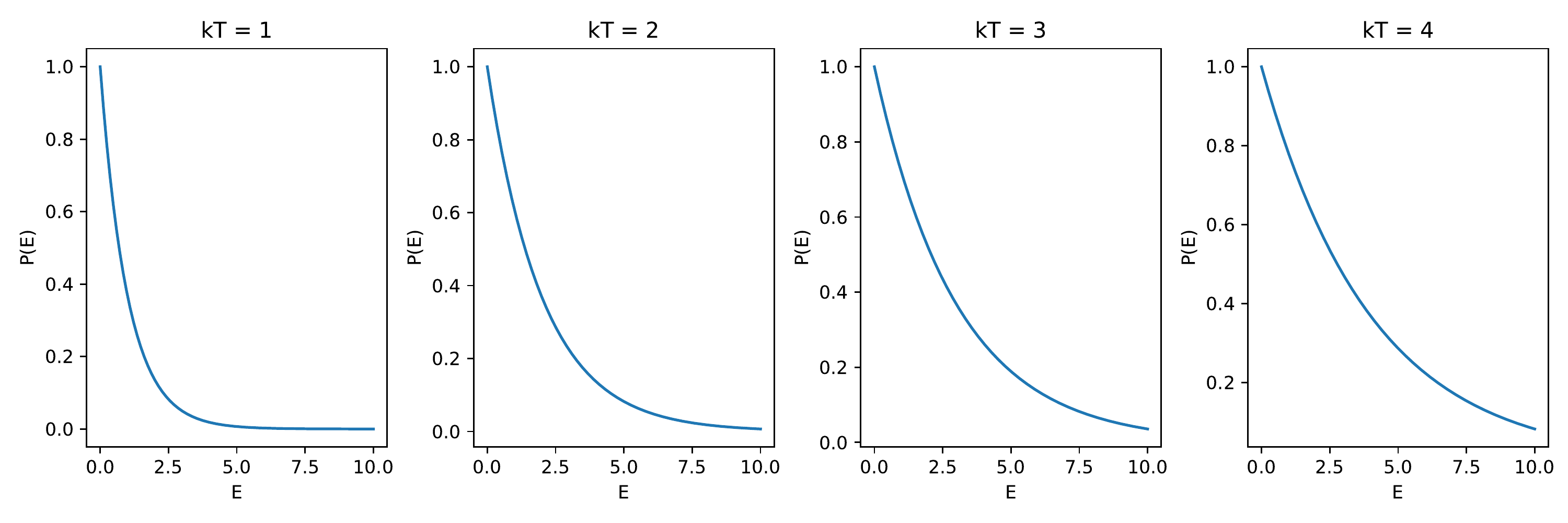}
    \caption{Graph of the Boltzmann distribution for various parameter values.}
    \label{fig:my_label6}
\end{figure}

The Sherrington-Kirkpatrick Model with an External Field, more commonly known as the \textit{Boltzmann machine}, introduced by Hinton and Sejnowski \cite{boltzmann1}, is a type of neural network architecture that uses a randomised approach to learning in what is termed \textit{stochastic learning}. This is commonly used for unsupervised ML tasks, such as pattern recognition, dimensionality reduction, and data generation \cite{boltzmann3}. Of particular importance is pattern recognition since DNA sequences exhibit recurring patterns, and it would prove advantageous to take skillfully make use of this feature. Building upon the ideas of the Boltzmann (or Gibbs) distribution from Statistical Mechanics \cite{boltzmann2}, which has the form
\begin{equation}
p(E_{i})=\frac{1}{Z}\times \exp\left(-\frac{E_{i}}{k_{B}T}\right),\quad\quad Z=\sum_{j=1}^{n}\exp\left(-\frac{E_{j}}{k_{B}T}\right), \tag{11}\label{11}
\end{equation}
where, in physical terms, $p$ is the probability, $E$ denotes the energy of the $i^{\text{th}}$ particle; $k_{B}\approx 1.3806\times 10^{-23}\;\text{J/K}$ is the Boltzmann constant, $T$ is the temperature, $j$ is a running dummy index, and $Z$ is known as the \textit{partition function}, the \textbf{Figure} \ref{fig:my_label6} shows this distribution for various parameter values. Of particular importance in ML is \textit{sampling}, \textit{videlicet} drawing information about the population based upon a sample. 

\begin{figure}[H]
    \centering
    \hspace*{-1.4cm}
    \includegraphics[scale=0.7]{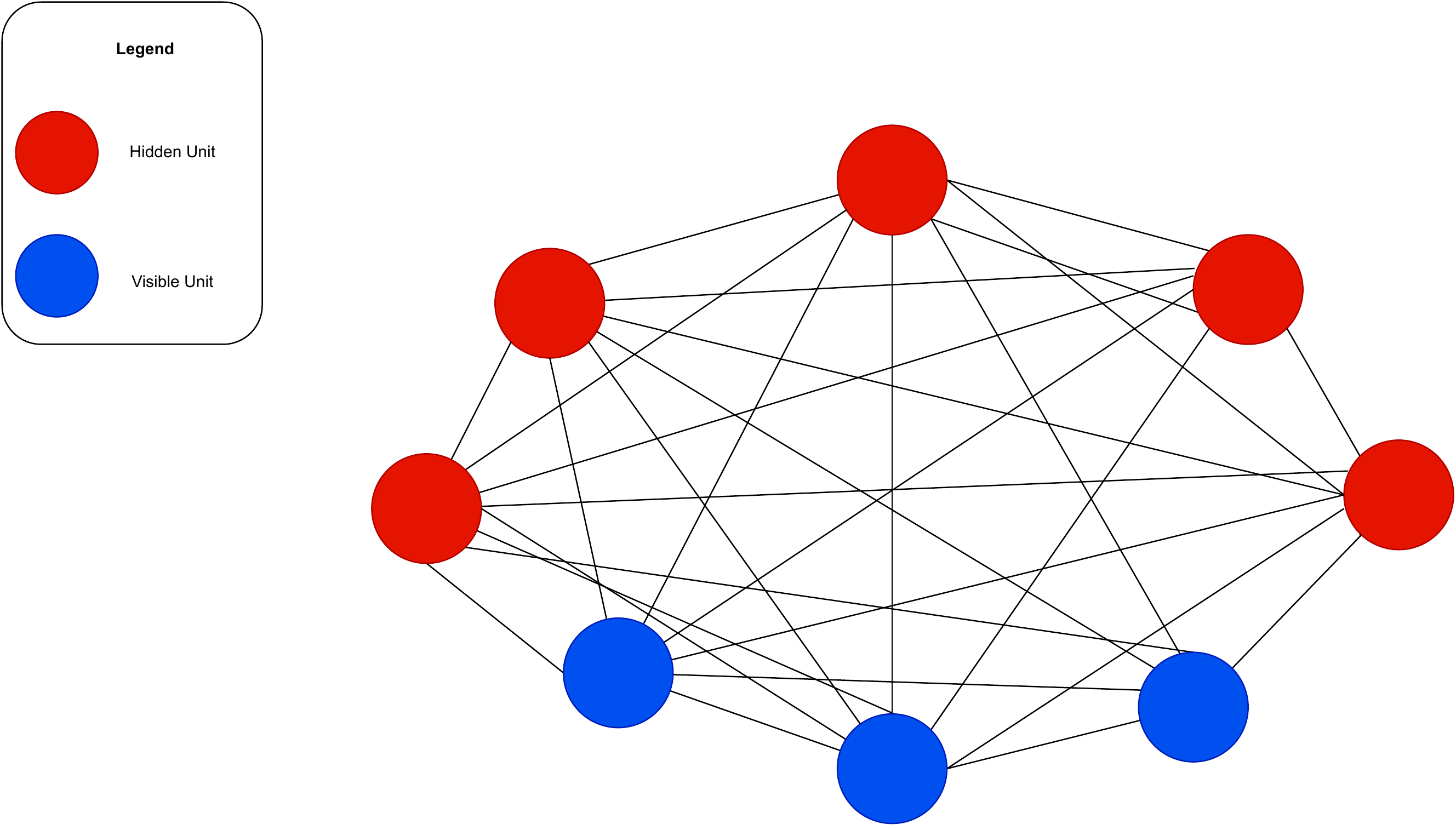}
    \caption{Architecture of a fully connected Boltzmann machine with $3$ visible units and $5$ hidden units.}
    \label{fig:my_label3}
\end{figure}

\vspace{-2pt}
\indent Geometrically, the Boltzmann machine consists of interconnected nodes / units called visible and hidden units representing input and output variables as illustrated in \textbf{Figure} \ref{fig:my_label3}. The connections between nodes are weighted and determine the probability of a certain unit state given the state of its neighbouring units. The visible units represent input data, whereas the Hidden units represent latent variables or features of the data. 
\\
\indent The machine is trained using Markov Chain Monte Carlo (MCMC) methods that simulate the Boltzmann distribution and allow the machine to find the optimal values for the weights that minimise the system's energy, representing the cost function of the ML system. This is achieved by sampling the Boltzmann distribution by constructing a Markov Chain with the desired distribution as its stationary distribution. Some standard MCMC techniques are: The Metropolis-Hastings algorithm, Gibbs Sampling, and Hamiltonian Monte Carlo \cite{boltzmann4}. The usage of MCMC is driven by the probability distribution being too difficult to sample from in the first place.

Two common architectural configurations of Boltzmann machines are
\begin{enumerate}
\item \textbf{Restricted Boltzmann Machine:} These are simplified versions of the Boltzmann machine, where the visible and hidden layers are connected in a bipartite graph. Both models, the Boltzmann machine and the restricted Boltzmann machine, use the same energy function to compute the joint probability of the visible and hidden nodes / layers (Eq. \ref{11}). 
\item \textbf{Deep Belief Networks:} These are a type of generative ML model consisting of multiple restricted Boltzmann machines (RBMs) stacked on top of each other. The hidden layer activation of one RBM serves as the visible layer activation of the next RBM to facilitate hierarchical representations of the data. They are mathematically analogous to the standard Boltzmann machine; however, they have two additional steps: Pre-training, whereby the weights and biases are initialised in the stack, and  fine-tuning, whereby the backpropagation algorithm is used to optimise the weights in the stack. 
\end{enumerate}

\indent On the other hand, we have seen recent success with \textit{Quantum Boltzmann Machines} (QBMs) \cite{boltzmann5}, a type of quantum neural network (QNN) that has been proposed as a potential way to use the power of QC for ML tasks. They are inspired by classical Boltzmann machines. In a QBM, the binary inputs are replaced with qubits connected by a network of quantum gates, which are used to implement the probabilistic updates that drive the learning process.
\\
\indent Using Boltzmann machines for classical-to-quantum data encoding, especially in DNA sequencing, is justifiable because the Boltzmann machine can learn pattern recognition and other structures in the input DNA sequence. Secondly, the Boltzmann machine can learn in an unstructured manner without requiring explicit labelling of nitrogenous bases and thus proffers a promising approach to leveraging both classical and quantum computing simultaneously. QBMs are used to model the joint probability distribution of quantum data, such as the state of a quantum system or the outcomes of quantum measurements. By learning this distribution, they can be used for various QML tasks, such as quantum tomography, quantum generative modelling, and quantum data compression.
\\
\indent Below, we provide the \textbf{Algorithm \ref{Qoltz}}  for testing a classical-to-quantum encoded DNA via the Boltzmann machine.
\newpage
\begin{algorithm}[H]
\caption{\texttt{Qoltz}$(\mathcal{D})$}
\begin{algorithmic}
\State input a DNA sequence $\mathcal{D}$
\State initialise the quantum circuit with $n,\ell,m\in\mathbb{N}, 0\leq\alpha\leq 1$, for the number of qubits, number of layers in the Boltzmann machine, number of steps in the optimisation algorithm, and the learning rate respectively with weights $\mathbf{W}$ and biases $\boldsymbol{\mathcal{B}}$
\For {each nitrogenous base $b$ in the DNA sequence $\mathcal{D}$}
\State apply \texttt{bin\_encode}$(b)$ to encode the bases as binary numbers
\State calculate the number of segments, $N$, to split the DNA sequence into, amongst $K$ varieties
\While{$\text{card}(\mathcal{D})\neq 0$}
\begin{equation*}   
N=\left\{\begin{matrix}
\frac{\text{card}(\mathcal{D})}{K},  \quad \text{N is even} \\  \left\lfloor\frac{\text{card}(\mathcal{D})-1}{K}\right\rfloor,\quad b_{\text{not sequenced}}=\left[\text{card}(\mathcal{D})-1\right]\;\left(\text{mod}\;K\right), \quad\text{N is odd}
\end{matrix}\right.
\end{equation*}
\EndWhile
\For {Each nitrogenous base $b$ in $N$}
\State apply \texttt{quant\_encode}$(s)$ to map 
$0\longrightarrow\ket{0}, 1\longrightarrow\ket{1}$ to obtain the corresponding quantum states of each segment $\ket{\psi}$
\Repeat 
\State split the encoded states into training and validation sets
\State sample a mini-batch of encoded states
\State calculate the energy for each quantum state in the mini-batch
\begin{align*}
E(\ket{\psi})=&\;\sum_{\mathcal{L}=1}^{\ell}\sum_{\mathcal{N}=1}^{n}\left[w_{0}(\mathcal{L}, \mathcal{N})\ket{\psi_{\mathcal{N}}}\otimes\ket{\psi_{\mathcal{N}+1}}+w_{1}(\mathcal{L}, \mathcal{N})\ket{\psi_{\mathcal{N}}}\otimes\ket{\psi_{\mathcal{N}+1}}\right. \\
&\left.+w_{2}(\mathcal{L}, \mathcal{N})\ket{\psi_{\mathcal{N}}}\otimes\ket{\psi_{\mathcal{N}+1}}\right]-\sum_{\mathcal{N}=1}^{n}\mathcal{B}_{\mathcal{N}}\ket{\psi_{\mathcal{N}}}
\end{align*}
\State calculate the partition function
\begin{equation*}
    Z=\sum_{\text{all states}}\exp(-\ket{\psi})
\end{equation*}
\State calculate the cost function, say 
\begin{equation*}
J=-\frac{1}{\text{card}(\mathcal{D})}\sum_{\text{all states}}\log\left[\frac{1}{Z}\times\exp(-E(\psi))\right]
\end{equation*}
\algstore{testcont}
\end{algorithmic}
\label{Qoltz}
\end{algorithm}
\begin{algorithm}[H]
\ContinuedFloat
\caption{\texttt{Qoltz}$(\mathcal{D})$ - Part $2$}
\begin{algorithmic}
\algrestore{testcont}

\State apply an update rule for the weights for each state, say
\begin{equation*}
    w\longleftarrow w-\alpha\nabla_{w}J(W,\ket{\psi})
\end{equation*}
\Until{$J\longrightarrow 0$}
\EndFor
\EndFor
\State \textbf{return} encoded states $\ket{\psi}$ 
\end{algorithmic}
\label{Qoltz2}
\end{algorithm}
In this \textbf{Algorithm \ref{Qoltz}} above, we have used the negative log-likelihood cost function; however, this can be adjusted to any cost function, such as MSE. In addition, we have used the SGD update rule; however, this can be changed to AdaGrad or ADAM.
\\
\indent
For example, let us consider the DNA sequence  \textbf{AAGT}; we will use the \texttt{Qoltz} algorithm (\ref{Qoltz}) to encode this DNA sequence into quantum states.
First, we must initialise the quantum circuit with the desired number of qubits, layers, steps, and learning rate. We can set $n=4$, $\ell=2$, $m=100$, and $\alpha=0.01$. We also need to initialise the weights and biases, but we will assume they are already initialised for simplicity.
Next, we will encode each nitrogenous base in the DNA sequence as binary numbers using the \texttt{bin\_encode} function. The encoding for each base is as follows: \textbf{A} $\rightarrow 00$, \textbf{C} $\rightarrow 01$, \textbf{G} $\rightarrow 10$, \textbf{T} $\rightarrow 11$. 
\\
\indent Thus, the encoding for the DNA sequence \textbf{AAGT} is $00000111$; Next, we need to split the encoded string into segments to train and validate the model on different parts of the sequence; this provides more flexibility to learn patterns and features specific to each segment and avoids overfitting any part of the DNA sequence. 
\\
\indent We want to split the string into $K=2$ varieties; the string length is $8$, divisible by $2$. Therefore, we can split the string into two equal segments: $0000$ and $0111$. We can map each segment to a quantum state using the \texttt{quant\_encode} function; in other terms, we use two qubits to encode each base, then, we can encode the first segment as $\ket{\psi_{1}} = \ket{00}\otimes\ket{00}$ and the second segment as $\ket{\psi_{2}} = \ket{10}\otimes\ket{11}$. 
\\
\indent We can now use the quantum states as input to the quantum Boltzmann machine algorithm (\ref{Qoltz}) to learn the underlying distribution of the DNA sequence. The algorithm will use the quantum states as input to a quantum circuit with $n$ qubits, $\ell$ layers, and $m$ optimization steps. The algorithm will then use an update rule to adjust the weights and biases of the quantum circuit to minimise the cost function $J$. The resulting encoded states can be returned as the algorithm's output and tested using the same algorithm. 
\section{The Dataset}
\label{sec:section7}
\begin{figure}[ht] 
\centering 
\includegraphics[width=0.8\textwidth]{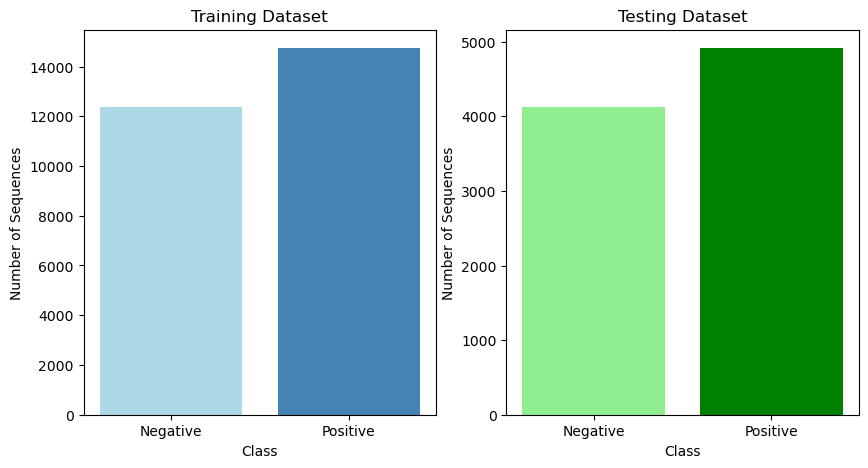} 
\caption{Training and testing data distributions.} \label{fig:dataset} 
\end{figure}

The dataset used in this work is called \texttt{human\_nontata\_promoters} \cite{gresova}; promoters generally are regions of DNA located upstream of a gene and serve as binding sites for RNA polymerase, an enzyme responsible for transcribing the gene into mRNA. The TATA box is a specific DNA sequence commonly found in the promoter region of genes and is essential for transcription initiation.
\newline
\indent Therefore, \texttt{human\_nontata\_promoters} refers to information related to promoter regions of human genes that do not contain the TATA box sequence. This data may include information about the location, function, and regulation of non-TATA promoters and their interactions with transcription factors and other regulatory elements. Understanding the characteristics of non-TATA promoters is essential for understanding the complexity of gene regulation in humans and identifying potential targets for therapeutic interventions.
\newline
\indent Our data consists of genomic intervals of length $251$. These intervals are located in the human genome's non--TATA promoter regions.
The dataset has two classes: positive and negative. These classes identify whether a given genomic interval contains an active promoter region (positive) or inactive (negative).
The dataset has a total of $36\;131$ sequences. Of these, $27\;097$ sequences are used for training, and $9\;034$ are used for testing.
The training set is further divided into two classes, negative and positive. The negative class contains $12\;355$ sequences, while the positive class contains $14\;742$ sequences.
The testing set is also divided into two classes, negative and positive. The negative class contains $4\;119$ sequences, while the positive class contains $4\;915$ sequences.
\newline
\indent There are five columns in this dataset
(\textbf{id}, \textbf{region}, \textbf{start}, \textbf{end}, \textbf{strand}), and each one has different information about the genomic intervals. Here is an examination of each column:
\begin{enumerate}
    \item \textbf{id:} This column contains a unique identifier for each genomic interval. This identifier is typically used to keep track of the individual sequences in the dataset.
    \item \textbf{region:} This column specifies the genomic region where the interval is located. In this case, the dataset includes only non-TATA promoter regions of the human genome.
    \item \textbf{start:} This column specifies the starting position of the genomic interval on the reference genome. The position is measured in base pairs (bp), the unit of length commonly used in genomics.
    \item \textbf{end:} This column specifies the ending position of the genomic interval on the reference genome. Like the start column, this position is also measured in base pairs.
    \item \textbf{strand:} This column specifies the orientation of the genomic interval to the reference genome. There are two possible orientations: forward $(+)$ or reverse $(-)$. The orientation of the interval is important because it determines the order of the nucleotides in the sequence.  
\end{enumerate}
Combining the information in these five columns makes uniquely identifying and locating each genomic interval in the dataset possible.
\newline
\indent This dataset is designed for a binary classification task to predict whether a given genomic sequence belongs to the ``positive'' or ``negative'' class. Using separate training and testing datasets for each class can help ensure that the resulting model can generalise well to new data. 
\newline 
\indent To illustrate the distribution of sequences in the training and testing datasets, we used a bar chart. \textbf{Figure} \ref{fig:dataset} shows the number of sequences in each class for the training and testing datasets. As can be seen from the figure, the training dataset contains more sequences than the testing dataset, and the number of positive sequences is higher than the number of negative sequences. This class imbalance can be challenging for some machine learning algorithms, and various techniques can be used to address it.
\newline 
\indent
To encode this type of data using quantum techniques, we first need to use the \textbf{id}, \textbf{region}, \textbf{start}, and \textbf{end} information to retrieve the genomic sequence using a genome assembly file or a genomic database. Once we have the sequence, we can encode it using quantum circuits, where a qubit represents each nucleotide base. By encoding the genomic sequence into qubits, we can exploit the quantum properties of entanglement and superposition to perform certain computations much faster than classical algorithms.
\newline
\indent Several quantum machine learning models can be applied to this dataset, such as quantum support vector machines, quantum neural networks, and quantum Boltzmann machines. These models can classify genomic sequences into positive or negative classes based on their underlying features. By using quantum techniques, we can potentially achieve faster and more accurate classification of genomic sequences, leading to significant advancements in personalised medicine and drug discovery.
\section{Conclusion and Future Work}
\label{sec:section8}
In this research, we have introduced novel algorithms that utilise various fields and advanced techniques from the field of QC to perform classical-to-quantum data encoding in the Bioinformatics domain. Our study provides an overview of existing classical-to-quantum encoding schemes and showcases how to use them in this field; For example, we demonstrate how the \texttt{Amplitude Encoding} method uses rotation gates to map classical data, in the form of vectors representing DNA sequences, to quantum states. Additionally, we have illustrated the use of \texttt{Pauli Feature Map Encoding}, which employs the \texttt{ZZFeatureMap} tool from Qiskit to obtain a \texttt{statevector} for the DNA sequence, and we employed the \texttt{IQPEmbedding} method from Pennylane to encode classical data into quantum states. Our results show that the resulting quantum states can be represented as vectors of complex numbers using IBM's \texttt{QasmSimulator}. We successfully demonstrated that data encoding could be applied to genomic data by leveraging these techniques.
\\
\indent The novelty of this paper was the introduction of new lossless compression algorithms, namely Quantum-influenced Huffman Coding (\texttt{QuantHuff}) and Quantum-influenced Burrows-Wheeler Transform (\texttt{QBWT}), which have been demonstrated to be effective using \texttt{Python} and Qiskit. We also introduced a wavelet-based encoding algorithm, \texttt{CosineEncoding}, that uses a two-dimensional Discrete Cosine Transform and quantum Fourier transform to map the magnitude of each coefficient to an $n$-qubit register, generating the encoded quantum states. We have also proposed the usage of information entropy using \texttt{SEncode} to encode the DNA sequences to quantum states based on the lowest entropy by  proposing two algorithms, \texttt{NZ22} and \texttt{NZ23}, that utilised reference data, and a third variety the used information geometry in the form of algorithm \texttt{QuantIG}. These algorithms offer various insights into the potential of quantum computing in Bioinformatics.
\\
\indent A point of contention that arises is that the algorithms for predictions, classifications, and comparisons are yet to be tested for their effectiveness. In so doing, we proposed one algorithm, \texttt{Qoltz}, based on energy methods and the Boltzmann machine to test the encoded data. Lastly, we discuss a potential dataset as a sandbox environment for testing against real-world scenarios. Furthermore, in \textbf{Table} \ref{table:encoding} in the appendix, we  comprehensively compare the proposed algorithms' advantages and disadvantages.
\\
\indent To further validate our results, we plan to investigate the performance of our proposed algorithms on larger datasets and compare them with existing methods, besides testing our algorithms on real quantum hardware, in follow-up studies and future iterations. Additionally, we will  explore the potential of quantum computing for other Bioinformatics tasks, such as protein folding prediction and drug discovery.
\\ 
\indent Moreover, we acknowledge the limitations and challenges of our proposed methods, such as the impact of noise on encoding and algorithm scalability. Additional research is needed to address these issues, and we hope to contribute to this ongoing effort. Altogether, this study offers a promising avenue for developing QC applications in Bioinformatics.


\section{Acknowledgements}
Firstly, the authors would like to extend their sincere gratitude and appreciation to the Quantumformalism.com community by the Zaiku Group for supporting us in undertaking this research project. Secondly, to Dr. Vesselin Gueorguiev, Dr. Dimitri Papaioannou, and Max Arnott for their insightful comments and astute suggestions with regard to the structure and content of this paper. 
\section{Code}
\label{code} 
All codes for the graphs and algorithms in the paper are available at: \url{https://github.com/quantumformalism/NISQ-C2QDE-GENOMICS}. 


\newpage
\section*{Appendix: Advantages and disadvantages of the quantum encoding algorithms}

\label{appendix:table}

\begin{table}[ht]
\caption{Advantages and disadvantages of our quantum encoding algorithms}
\centering
\begin{tabular}{|p{2cm}|p{7cm}|p{7cm}|}
\hline
\textbf{Algorithm} &\textbf{Advantages} &\textbf{Disadvantages} \\ 
\hline
\texttt{Amplitude Encoding} (\ref{amplitude}) & 
- Can handle high-dimensional data, as high-dimensional vectors can represent DNA sequences. \newline
- Provides flexibility in choosing rotation angles based on the input data. &
- Requires normalisation of input data. \newline
- Sensitivity to input data, as small changes in the DNA sequence can result in large changes in the output. \newline
- The encoding is not reversible, so decoding the state may be complex. \\ \hline
\texttt{Pauli Feature Map Encoding} (\ref{pauli}) &  
- Can handle high-dimensional classical data points. \newline
- Can be customised based on the application. &
- Requires hardware that can handle controlled Z gates and nonlinear functions. \\ \hline
\texttt{QuantHuff} (\ref{QuantHuff}) & 
- Provides a compression technique that assigns variable-length binary codes to each symbol in a given set of symbols. \newline 
- Can reduce the amount of memory required to store the data. &
- Requires classical operations to construct a Huffman tree and assign binary codes. \\ \hline
\texttt{QBWT} (\ref{agbwt}) & 
- Can be used for long DNA sequences. & 
- Requires hardware that can handle unitary transformations. \\ \hline
\texttt{Cosine Encoding}(\ref{algi}) & 
- Relatively simple to implement compared to other quantum encoding methods. &
- Requires many qubits, making it difficult to scale to larger images. \newline
- May require a high degree of precision. \\ \hline 

\end{tabular}
\label{table:encoding}
\end{table}
\begin{table}[ht]
\centering
\begin{tabular}{|p{2cm}|p{7cm}|p{7cm}|}
\hline
\texttt{SEncode} (\ref{SEncode}) & 
- Can encode long DNA sequences using relatively few qubits. \newline
- Using a unitary operator to map each quantum state to a computational basis state ensures that the encoding is reversible. &
- The performance may be sensitive to the choice of segment size and the number of segments. \newline
- It may require significant resources, including qubits and classical computational power. \\ \hline
\texttt{NZ22} (\ref{NZ22}) & 
- It considers the probability distributions of both the input and reference DNA sequences, which may lead to more accurate encoding compared to algorithms that only consider one sequence.\newline
- The number of qubits required for encoding is determined based on the KL divergence, which measures the difference between the probability distributions. This may lead to more efficient encoding than the other algorithms using a fixed number of qubits. &
- The mapping of each base to an $n$-qubit register may result in many qubits required for longer sequences, which may lead to scalability issues. \newline
- The calculation of the KL divergence can be computationally intensive, especially for longer sequences, which may increase the time required for encoding. \\ \hline
\texttt{NZ23} (\ref{NZ23}) & 
- It is based on the Bhattacharyya divergence, a robust similarity measure between probability distributions. \newline
- The number of qubits required for encoding the state is determined based on the similarity between the input and reference sequences. This can lead to more efficient encoding than other methods using a fixed number of qubits. &
- It does not consider the specific features of the DNA sequence, such as the presence of repeats or structural variations, which can affect the accuracy of the encoding. \newline
- The adjustable hyperparameter $\alpha$ needs to be tuned for optimal performance, which may require trial and error. \\ \hline

\hline
\end{tabular}
\end{table}
\begin{table}[ht]
\centering
\begin{tabular}{|p{2cm}|p{7cm}|p{7cm}|}
\hline
\texttt{QuantIG} (\ref{quantig}) & 
- Using a Riemannian manifold to map probability distributions to a Hilbert space allows for efficient and accurate quantum operations on the probability distributions. \newline
- It considers the distance between the two probability distributions using an information metric, which can provide more accurate results than other algorithms considering only probability distributions. &
- Using a Riemannian manifold and state mapping may require more computational resources than simpler algorithms. \newline
- Using a metric operator on the Hilbert space may require significant mathematical expertise, limiting accessibility. \newline
- The accuracy and efficiency may depend on the choice of information metric and other parameters, which could require further experimentation and optimisation. \\ \hline
\texttt{Qoltz} (\ref{Qoltz}) & 
- The quantum approach may offer faster processing times and handle larger datasets. \newline
- The ability to perform quantum operations on the encoded states could lead to new insights and discoveries in DNA research. &
- The effectiveness is not yet established compared to classical algorithms, and comparing their performance may not be easy. \newline
- The optimisation process may be computationally intensive and require significant resources and time. \\ \hline
\end{tabular}
\end{table}

\end{document}